\documentclass[usenatbib]{mnras}

\usepackage{subfigure}
\usepackage{graphicx}	% Including figure files
\usepackage{amsmath}	% Advanced maths commands
\usepackage{amssymb}	% Extra maths symbols
\usepackage{multicol}        % Multi-column entries in tables
\usepackage{bm}		% Bold maths symbols, including upright Greek
\usepackage{pdflscape}	% Landscape pages
\usepackage{color}
\usepackage{blindtext}
\title[]{Globular clusters as a probe for Weyl Conformal Gravity}
\author[Tousif Islam]{Tousif Islam$^{1}$\thanks{Email : tousifislam24@gmail.com},\\
	$^1$ International Centre for Theoretical Sciences, Tata Institute of Fundamental Research, Bangalore- 560012, India}

\pagerange{\pageref{firstpage}--\pageref{lastpage}} \pubyear{2019}
\def\LaTeX{L\kern-.36em\raise.3ex\hbox{a}\kern-.15em
	T\kern-.1667em\lower.7ex\hbox{E}\kern-.125emX}

\begin{document}
	\label{firstpage}
	\pagerange{\pageref{firstpage}--\pageref{lastpage}}

\maketitle

%%%%%%%%%%%%%%%%%%%%%%%%%%%%%%%%%%%%%%%%%%%%%%%%%%%%%%%%%%%%%%%%%%%%%%%%%%%%%%%%%%%%%%%%%%%%%%%%
	
\begin{abstract}
Eventual flattening of velocity dispersion profiles of some galactic globular clusters in the Milky Way cannot be explained in the framework of Newtonian gravity and hence in general theory of relativity in the weak field limit, without resorting to the occurrence of tidal effects. We explore the possibility of explaining such deviation from expected Keplerian fall-off in dispersion profiles within the context of Weyl conformal gravity. We choose a set of 20 globular clusters for which recent kinematic measurements are available. We model the globular clusters with approximate Hernquist mass profiles and choose a constant mass-to-light ratio throughout the cluster as the only free parameter in the model. Our analysis finds reasonable Weyl gravity fits to the observed dispersion profiles, exhibiting both Keplerian decline and eventual flattening, with acceptable mass-to-light ratios.  We further recover a Tully-Fisher like scaling relation in globular clusters through Weyl gravity.
\end{abstract}	

\begin{keywords}
gravitation -- stars: kinematics and dynamics -- globular clusters: general -- Galaxy: kinematics and dynamics.
\end{keywords}
%%%%%%%%%%%%%%%%%%%%%%%%%%%%%%%%%%%%%%%%%%%%%%%%%%%%%%%%%%%%%%%%%%%%%%%%%%%%%%%%%%%%%%%%%%%%%%%%
%%%%%%%%%%%%%%%%%%%%%%%%%%%%%%%%%%%%%%%%%%%%%%%%%%%%%%%%%%%%%%%%%%%%%%%%%%%%%%%%%%%%%%%%%%%%%%%%
%%%%%%%%%%%%%%%%%%%%%%%%%%%%%%%%%%%%%%%%%%%%%%%%%%%%%%%%%%%%%%%%%%%%%%%%%%%%%%%%%%%%%%%%%%%%%%%%
%%%%%%%%%%%%%%%%%%%%%%%%%%%%%%%%%%%%%%%%%%%%%%%%%%%%%%%%%%%%%%%%%%%%%%%%%%%%%%%%%%%%%%%%%%%%%%%%
\section{Introduction}
Weyl conformal gravity has originally been proposed by \cite{weyl1918} and has later been re-studied by \cite{weyl1}. The major motivations to pursue Weyl  gravity have been  to explain the apparent `mass discrpancies' in galaxies and clusters, and the observed cosmic speed-up of the universe without the ad-hoc addition of exotic Dark Matter \citep{bertone2005particle} and Dark Energy \citep{peebles2003cosmological} respectively. Like any other alternative or modified gravity theory, Weyl gravity tries to achieve these through replacing Einstein's General Relativity (GR) with new laws of gravity.  The theory has been able to generate enough interest due to its renormalizability, embedded conformal symmetry and the absence of ghosts \citep{bender2008no}. \\

Weyl gravity has been successfully used to fit the rotation curves for several galaxies without resorting to the dark matter \citep{weylrot2,weylrot1,weylrot3,weylrot4}. Unlike other modified gravity theories, Weyl gravity predicts an eventual decline of the rotation curve after the flat portion for each and every galaxy at large enough distances from the galactic center. This effect could only be tested with galaxies for which rotation curve data extends well beyond the optical length. Recent studies with the extended rotational velocity data for the Milky Way (MW) have found reasonable agreement between the prediction and the observed profile  \citep{obrien,kt2018}. However, the construction of the MW rotation curve is heavily influenced by the uncertainties in measurements of the anisotropy parameter, and the velocity and Galactocentric distance of the Sun. \cite{kt2018} (DI18) have further demonstrated that the success of Weyl gravity to account for the MW rotation curve is robust against these current uncertainties. The theory is also found to be consistent with solar system phenomenology \citep{perihelion,mannheim2007schwarzschild}. \cite{weylrot5} has further showed that Weyl gravity agrees with supernova data for redshift $z \sim$1. However, its ability to account for observations at the scale of galaxy clusters remains inconclusive \citep{kt2018,weylcluster1,weylcluster2,weylcluster3}. It is also currently unclear whether Weyl gravity can fit the merging dynamics of the Bullet clusters  \citep{clowe2006direct}, which poses a major challenge to any modified gravity theory.\\

From astrophysical point of view, further tests for Weyl gravity could be formulated using the observed velocity dispersion profiles of the galactic globular clusters (GCs) in the MW. GCs are generally thought to be devoid of dark matter \citep{gcdm1,gcdm2}. One should therefore expect the velocity dispersion of GCs to follow a Keplerian fall-off and vanish at large distances from the center of the cluster given GR is valid. However, observed dispersion profile for some GCs exhibit a different trend. The dispersion is found to be maximum at the center and then it gradually decreases before it settles down for an asymptotic value \citep{scarpa2003using,scarpa2004using,ngcothers2,scarpa2007using}. One possible explanation for this puzzling observation could be tidal heating \citep{drukier2007global,kupper2010,kennedy2014application}. However,  \cite{gctidal} have not found any convincing evidence for this hypothesis. An alternative explanation could be a modification of the gravitation law effective on this length-scale  \citep{moffattoth,haghi2009testing,haghi2011distant}. The apparent similarity between the flat dispersion profiles of the elliptical galaxies and galactic GCs has further bolstered the idea.\\

DI18 have already explored whether Weyl gravity can account for the observed flattening of dispersion profiles for a set of four GCs (NGC 288, NGC 1851, NGC 1904 \& NGC 5139) which have different luminosities, sizes and dynamical histories. However, the data used in their study had been obtained from \cite{scarpa2004using,ngcothers2,scarpa2007using} which has recently been questioned and superseded by larger and more accurate datasets. In this paper, we extend the analysis to a larger set of GCs with updated velocity dispersion data. Current radial velocity dispersion data of these GCs have been compiled by combining the proper motion dispersion data from \textit{Gaia} DR2 and stellar line-of-sight velocities \citep{watkins2015hubble,baumgardt2016n,kamann2017stellar,baumgardt2018mean,baumgardt2018catalogue}.  Our sample includes 7 GCs for which \cite{ngc1851a1904} claimed to observe an eventual flattening of dispersion profile that they argued is due to a modification of the laws of gravity at those length-scales. Furthermore, we choose another 13 GCs for which excellent kinematic data is available. Interestingly, many of these GCs show no sign of flattening in the velocity dispersion data. Therefore, combined with the first set of 7 GCs, they provide an excellent tool to test any alternative theory of gravity. \\

The paper is organized in the following way. In Section \ref{sec1}, we would briefly introduce Weyl gravity. The task of testing Weyl gravity against velocity dispersion of GCs will then be taken up in Section \ref{sec2}. We would present the mass model used for GCs in Section \ref{sec2a}. The theoretical formulation of the velocity dispersion in the context of Weyl gravity will be developed in Section \ref{sec2b}. Subsequently, we would fit the observational data and present our results in Section \ref{sec2c}. Finally, we would discuss the implication of our result and conclude in Section \ref{sec3}.

\section{Weyl Conformal Gravity}
\label{sec1}
Weyl conformal gravity (see \cite{weylrot5} for a detailed review) employs the principle of local conformal invariance of the space-time under the transformation $g_{\mu \nu} (x) \rightarrow \Omega^{2}(x) g_{\mu \nu} (x)$, where  $g_{\mu \nu}$ is the metric tensor and $\Omega(x)$ is a smooth strictly positive function. Such requirement leads to a unique scalar action
\begin{align}
I_{w}=-\alpha_{g} \int d^{4}x \sqrt{-g} C_{\lambda\mu\nu\kappa} C^{\lambda\mu\nu\kappa},
\label{eqn1a}
\end{align}
where $\alpha_{g}$ is a dimensionless coupling constant and $C_{\lambda\mu\nu\kappa}$ is the Weyl tensor \citep{weyl1918} which is expressed as a combination of
the Riemann tensors $R_{\lambda\mu\nu\kappa}$, Ricci tensors $R_{\mu\kappa}$ and the Ricci scalar $R^{\nu}_{\nu}$:
\begin{align}
& 
\begin{aligned}[t]
C_{\lambda\mu\nu\kappa}= & R_{\lambda\mu\nu\kappa} - \frac{1}{2}(g_{\lambda\nu}R_{\mu\kappa} 
- g_{\lambda\kappa}R_{\mu\nu} - g_{\mu\nu}R_{\lambda\kappa} + g_{\mu\kappa}R_{\lambda\nu} ) \\
&+ \frac{1}{6}   R^{\alpha}_{\alpha}( g_{\lambda\nu}g_{\mu\kappa} - g_{\lambda\kappa}g_{\mu\nu} ).
\end{aligned}
\label{eqn1b}
\end{align}
The action then reduces to:
\begin{align}
&
\begin{aligned}[t]
I_{w}= -2 \alpha_{g} \int d^{4}x \sqrt{-g} [ R_{\lambda\mu\nu\kappa} R^{\lambda\mu\nu\kappa}- 2 R_{\mu\kappa}R^{\mu\kappa} + \frac{(R^{\nu}_{\nu})^{2}}{3} ] ,
\end{aligned}
\label{eqn1}
\end{align}
This action leads to the following fourth order field equation instead of the usual second order field equation in GR \citep{weyl1}:
\begin{eqnarray}
4 \alpha_{g} W^{\mu\nu} = 4 \alpha_{g} (C_{;\lambda;\kappa}^{\lambda\mu\nu\kappa} - \frac{1}{2} R_{\lambda\kappa} C^{\lambda\mu\nu\kappa} ) = T^{\mu\nu} , 
\label{eqn2}
\end{eqnarray}
where $W^{\mu\nu}$ is the Bach tensor, $T^{\mu\nu}$ is the matter-energy tensor and   `;' denotes covariant derivative. The non-linear nature of the field equation makes it difficult to obtain analytical solutions. However, for static spherically symmetric geometry, the line element can be written as \citep{weyl1}
\begin{equation}
ds^2= \left[-B(r) dt^2+ {dr^2\over B(r)} + r^2
d\Omega_2\right]~~.
\label{eqn3}
\end{equation}
The field equation then reduces to a simpler fourth order Poisson equation in the weak field limit \citep{weylrot5}
\begin{equation}                                                                               
\nabla^4 B(r) =  f(r) ~~,
\label{eqn4}
\end{equation}
where $f(r)$ is the source function. The general solution then immediately reads
\begin{eqnarray}
B(r)=  1 - \frac{2\beta}{r} +\gamma r - \kappa r^{2},
\label{eqn5}
\end{eqnarray}
with $\gamma = $ $-\frac{1}{2}$ $\int^{r}_{0}dr^{'}r^{'2}f(r^{'})$ ; $2\beta= \frac{1}{6} \int^{r}_{0}dr^{'}r^{'4}f(r^{'})$; and $\kappa = \frac{r^{2}}{6} \int^{\infty}_{r}dr^{'}r^{'}f(r^{'})$. It could thus be concluded that $\beta$ and $\gamma$ originates completely from the local mass distribution , and  $\kappa$ has a global origin. Therefore, in conformal gravity, both local matter inside the source as well as matter exterior to it contributes to the solution of Eq. \ref{eqn4} \citep{weylrot5}. Identifying $B(r)=1 + \frac{2 \phi}{c^{2}}$, with $c$ being the speed of light, one may write the effective potential for  a point source to be
\begin{eqnarray}
\frac{ \phi}{c^{2}} =   - \frac{\beta}{r} +\frac{\gamma r}{2} - \frac{\kappa}{2} r^{2}.
\label{eqn6}
\end{eqnarray}
We note that in addition to the Newtonian term, the solution features a linear potential term, important on galactic scales, and a quadratic term, important on cosmological scales. Successful fitting to galaxy rotation curves further requires $\gamma$ to be broken into two parts: $\gamma=\gamma_{0} + \Big( \frac{M}{M_{\odot}}\Big) \gamma^{*}$, where $M$ is the point-source mass \citep{weylrot2}. The potential thus becomes 
\begin{eqnarray}
\frac{ \phi}{c^{2}} =   - \Big( \frac{M}{M_{\odot}}\Big)\frac{\beta^{*}}{r}  +\Big( \frac{M}{M_{\odot}}\Big)\frac{\gamma^{*} r}{2} +\frac{\gamma_{0} r}{2} - \frac{\kappa}{2}  r^{2},
\label{eqn7}
\end{eqnarray}
where $\beta =\Big( \frac{M}{M_{\odot}}\Big) \beta^{*}$. Weyl gravity thus possess four universal parameters: $\beta^{*}$, $\gamma^{\ast}$,  $\gamma_0$ and $\kappa$. Previous fits to galaxy rotation curves \citep{weylrot2,weylrot1,weylrot3,weylrot4} yielded the following values for the Weyl gravity parameters: $\beta^{\ast} = 1.48 \times 10^{5}$ $cm$; $\gamma^{\ast} = 5.42 \times 10^{-41}$ $cm^{-1}$; $\gamma_0 = 3.06 \times 10^{-30}$ $cm^{-1}$ and $\kappa = 9.54 \times 10^{-54} $ $cm^{-2}$. However, a more convenient parameterization for the potential would be \citep{weylcluster2}
\begin{equation}
 \phi= -\frac{GM}{r} + \frac{GM}{R^{2}_{0}}r + \frac{GM_{0}}{R^{2}_{0}}r - \frac{\kappa}{2}  r^{2}c^{2} ,
\label{eqn8}
\end{equation}
where the $\gamma_{0}$ and $\gamma^{*}$ translates to $R_0=\Big[ \frac{2GM_\odot}{\gamma_*c^2} \Big]^{1/2}=24$~kpc and $M_0=\frac{\gamma_0}{\gamma^*} M_\odot = 5.6\times 10^{10}M_\odot$. The third term generates a constant acceleration $\frac{GM_{0}}{R^{2}_{0}}$ independent of the local source and is attributed to the homogeneous cosmological background \citep{weylrot5}. The fourth term, on the other hand, incorporates the effect of inhomogeneities in the cosmological background \citep{weylrot5}. Therefore, in Weyl gravity, the potential around a point mass is a summation of contribution from both local and global effects. 
%%%%%%%%%%%%%%%%%%%%%%%%%%%%%%%%%%%%%%%%%%%%%%%%%%%%%%%%%%%%%%%%%%%%%%%%%%%%%%%%%%%%%%%%%%%%%%%%
%%%%%%%%%%%%%%%%%%%%%%%%%%%%%%%%%%%%%%%%%%%%%%%%%%%%%%%%%%%%%%%%%%%%%%%%%%%%%%%%%%%%%%%%%%%%%%%%
%%%%%%%%%%%%%%%%%%%%%%%%%%%%%%%%%%%%%%%%%%%%%%%%%%%%%%%%%%%%%%%%%%%%%%%%%%%%%%%%%%%%%%%%%%%%%%%%
%%%%%%%%%%%%%%%%%%%%%%%%%%%%%%%%%%%%%%%%%%%%%%%%%%%%%%%%%%%%%%%%%%%%%%%%%%%%%%%%%%%%%%%%%%%%%%%%
\section{Testing Weyl gravity with Globular clusters}
\label{sec2}
%%%%%%%%%%%%%%%%%%%%%%%%%%%%%%%%%%%%%%%%%%%%%%%%%%%%%%%%%%%%%%%%%%%%%%%%%%%%%%%%%%%%%%%%%%%%%%%%
%%%%%%%%%%%%%%%%%%%%%%%%%%%%%%%%%%%%%%%%%%%%%%%%%%%%%%%%%%%%%%%%%%%%%%%%%%%%%%%%%%%%%%%%%%%%%%%%
\subsection{Mass profile of GCs}
\label{sec2a}
We consider simple approximate models for the GCs. We assume GCs to be spherically symmetric and non-rotating. The  mass-to-light ratio of GCs, $\frac{M}{L}$, generally exhibits a radial variation \citep{lane2010}. However, for the sake of simplicity, we take it to be constant throughout the cluster.
We model the mass distribution of the cluster using a simple Hernquist profile \citep{hern}
\begin{equation}
\rho_\mathrm{hern}(r)=\frac{Mr_0}{2\pi r(r+r_0)^3},
\label{eqn9}
\end{equation}
where $M=(\frac{M}{L}) L$ is the total mass of the cluster, and $r_0$ is a characteristic radius.  $L$ denotes the total luminosity of the cluster. The characteristic radius in Hernquist profile is related to the half-light radius through $r_h=1.8153 r_0$  \citep{hern}. Total luminosity of individual GCs has been inferred from the reported total mass and mass-to-light ratio of GCs in \cite{baumgardt2018catalogue} obtained through fitting a large set of N-body simulations to their velocity dispersion and surface density profiles. We also use the half-light radius reported in \cite{baumgardt2018catalogue} to compute the characteristic radius $r_0$. We list these values in Table \ref{tab1}. We note that one could choose several other mass profiles, such as King's profile \citep{king1966structure} or Plummer profile \citep{plummer1911problem}, to model GCs more accurately. However, we find that the choice of a different model does not alter our final conclusion much. We address this issue in Section \ref{sec2c}.
%5++++++++++++++++++++++++++
\begin{table}
\caption{Mass distribution of GCs:  Half-light radius, luminosities and cluster distances for different GCs used in this paper are listed here \citep{baumgardt2018catalogue}. }
\label{tab1}
\begin{center}
\begin{tabular}{lccc}
\hline
GC & half-light radius & Luminosity &Distance\\
& $pc$ & $L_{\sun}$ &$kpc$\\
\hline
NGC 104 &3.57 &4.401 $\times 10^{5}$ &4.41\\
NGC 288 &6.99 &4.85 $\times 10^{4}$ &9.80\\
NGC 362 &2.32 &2.09 $\times 10^{5}$ &9.40\\
NGC 1851 &1.65 &1.49 $\times 10^{5}$ &11.40\\
NGC 1904 &2.59 &9.83 $\times 10^{4}$ &13.27\\
NGC 2808 &2.06 &4.52 $\times 10^{5}$ &9.80\\
NGC 5139 &7.04 &1.22 $\times 10^{6}$ &5.20\\
NGC 5904 &3.61 &2.45 $\times 10^{5}$ &7.50\\
NGC 5927 &4.98 &1.36 $\times 10^{5}$ &8.40\\
NGC 6171 &3.11 &4.03 $\times 10^{4}$ &6.09\\
NGC 6266 &1.83 &2.75 $\times 10^{5}$ &6.47\\
NGC 6341 &2.28 &1.48 $\times 10^{5}$ &8.10\\
NGC 6362 &5.77 &5.65 $\times 10^{4}$ &8.00\\
NGC 6388 &1.96 &5.49 $\times 10^{5}$ &11.00\\
NGC 6397 &2.19 &4.08 $\times 10^{4}$ &2.48\\
NGC 6411 &2.03 &6.00 $\times 10^{5}$ &12.00\\
NGC 6656 &3.26 &1.93 $\times 10^{5}$ &3.10\\
NGC 7078 &1.90 &3.39 $\times 10^{5}$ &9.90\\
NGC 7089 &3.00 &3.59 $\times 10^{5}$ &11.50\\
NGC 7099 &2.44 &7.19 $\times 10^{4}$ &8.10\\
\hline
\end{tabular}
\end{center}
\end{table}
%5++++++++++++++++++++++++++

%%%%%%%%%%%%%%%%%%%%%%%%%%%%%%%%%%%%%%%%%%%%%%%%%%%%%%%%%%%%%%%%%%%%%%%%%%%%%%%%%%%%%%%%%%%%%%%%
%%%%%%%%%%%%%%%%%%%%%%%%%%%%%%%%%%%%%%%%%%%%%%%%%%%%%%%%%%%%%%%%%%%%%%%%%%%%%%%%%%%%%%%%%%%%%%%%
\subsection{Velocity dispersion}
\label{sec2b}
The gravitational potential of a point source has been given by Eq. (\ref{eqn8}). We have already pointed out that the first two terms originate from local source while the last two are global terms. Thus, for an extended source, the first two terms will be modified while the last two would remain unaltered \citep{weylrot5}.\\

To compute the gravitational potential within GCs, we assume them to be self-gravitating spheres. We adopt the formalism outlined by \cite{weylcluster2} and  \cite{weylcluster1}. We first consider a homogeneous spherical shell of density $\rho$, radius $R$ and mass $m=4\pi \rho R^2 {\rm d}R$.  Initially, we ignore the global terms and only concentrate on the contributions from the local sources. The total potential of the shell then reads \citep{weylcluster1}
\begin{equation}
\phi_{sh}(r) = G m \left\{\begin{array}{ll}
-{1\over R} + {1\over R_0^2}\left({r^2\over 3R} + R\right) & r<R\\
-{1\over r} + {1\over R_0^2}\left({R^2\over 3r} + r\right) & r>R \; . \\
\end{array}\right.
\end{equation}
The local contribution to the gravitational potential of a self-gravitating sphere is then obtained as
\begin{eqnarray}
{\phi(r)\over G}=-{I_0(r)\over r} & -& E_{-1}(r) +  {1\over R_0^2}\left[{I_2(r)\over 3r} 
+ rI_0(r)\right.\cr 
& + & \left. {r^2\over 3}E_{-1}(r) + E_1(r)\right] ,
\label{eqn10}
\end{eqnarray} 
where the interior and exterior moments of the mass is defined respectively as 
\begin{equation}
I_n(r)=4\pi \int_0^r \rho(x)x^{n+2} dx,
\end{equation}
and
\begin{equation}
E_n(r)=4\pi \int_r^{+\infty} \rho(x)x^{n+2}dx.
\end{equation} 
The inward gravitational acceleration arising from the local sources in a self-gravitating sphere is thus \citep{weylcluster1}

\begin{align}
&
\begin{aligned}[t]
&a_{local}(r) = -{\nabla\phi(r)}  \\
&= G  \left[-{I_0(r)\over r^2} + {1\over R_0^2}\left({I_2(r)\over 3 r^2} - {2\over 3} r E_{-1}(r) 
- I_0(r)\right) \right]\; .
\end{aligned}
\end{align}

\begin{figure*}
	\centering
	\includegraphics[width=0.9\linewidth]{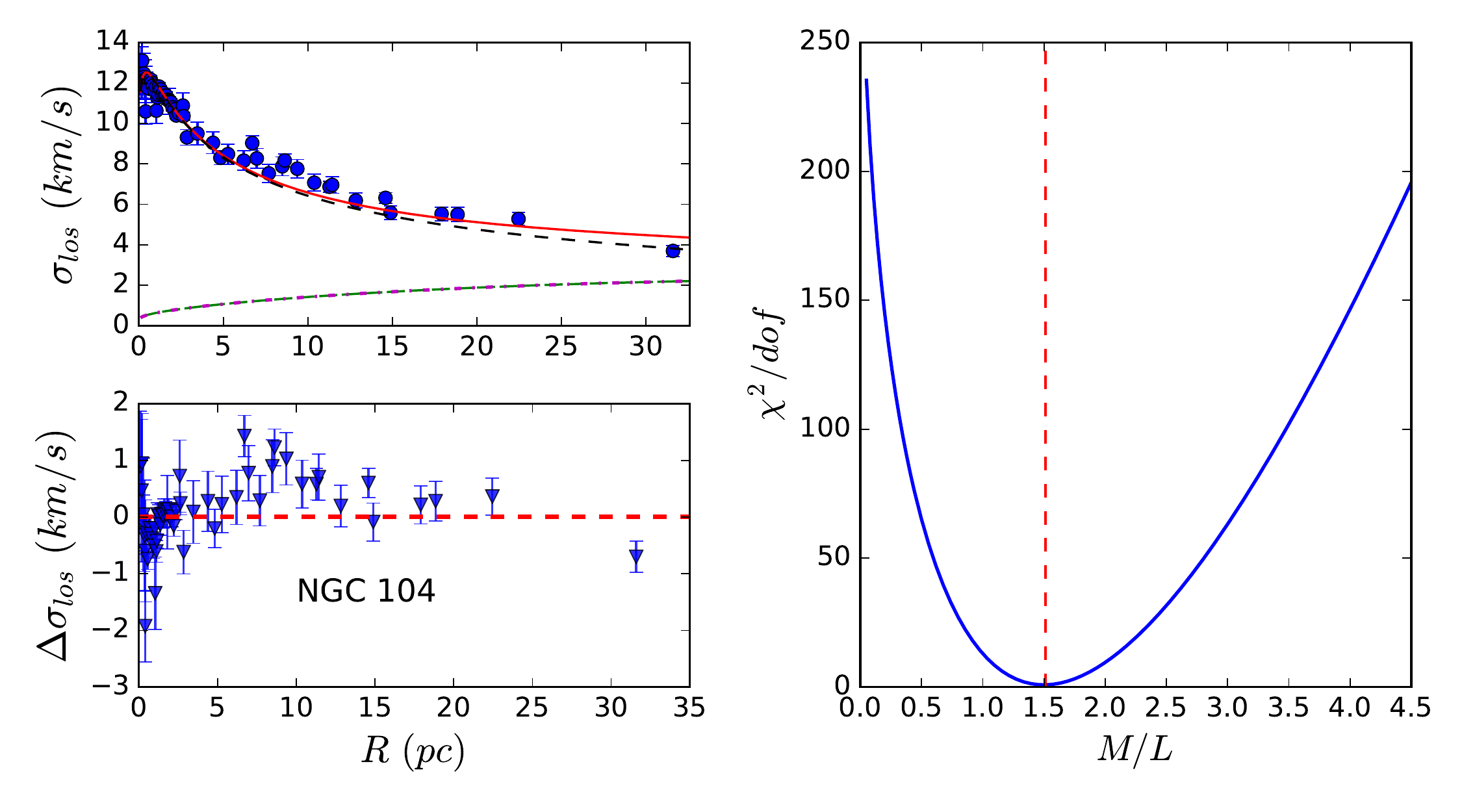}
	\caption[]{\textbf{\textit{In Left :} Velocity dispersion profile of the NGC 104 has been plotted as a function of projected radial distances (upper panel). Observed velocity dispersion is plotted in blue circles with errorbars and Weyl gravity best-fit is plotted as a solid red line. Additionally, we show the contribution due to the Newtonian term alone (first term in Eq. \ref{eqn8}) in black dashed line and the contribution from the two linear terms only (second and third term in Eq. \ref{eqn8})  in long dashed green and the contribution from the linear and quadratic terms together in dotted magenta. The residual profile is shown in the lower panel. \textit{In Right :} We plot the reduced chi-square profile as a function of mass-to-light ratio. The best-fit value for $\frac{M}{L}$, obtained through chi-square minimization, is shown as a vertical dashed line in red.}}
	\label{fig1}
\end{figure*}

\noindent Finally, we incorporate the effects of the global terms. The final expression for acceleration now reads
\begin{align}
&
\begin{aligned}[t]
&a(r)  \\
&= G  \left[-{I_0(r)\over r^2} + {1\over R_0^2}\left({I_2(r)\over 3 r^2} - {2\over 3} r E_{-1}(r) 
- I_0(r)\right) \right] \\
& + {GM_0\over R_0^2}  - \kappa c^2 r \; .
\end{aligned}
\end{align}
For spherically symmetric and non-rotating systems like GCs, the velocity dispersion is given by the Jeans equation \citep{bt1987}
\begin{equation}
\frac{\partial(\rho(r)\sigma^2(r))}{\partial r} +\frac{2 \rho(r) \xi \sigma^{2}(r)}{r} = \rho(r) a(r),
\label{eqn15}
\end{equation}
where $r$ is the radial distance from the GC center and $\rho(r)$ is the radial density distribution function. We now utilize the constraint $\lim\limits_{r\rightarrow\infty}\rho(r)\sigma^2(r)=0$. Additionally, we assume anisotropy parameter $\xi=0$. Eq. (\ref{eqn15}) thus gives
\begin{equation}
\sigma^2(r)=\frac{1}{\rho(r)}\int\limits_r^\infty\rho(r')a(r')~dr'.
\label{eqn16}
\end{equation}
\noindent Finally, the corresponding  projected line-of-sight (LOS) velocity dispersion reads   [see Eq. (14-16) in \cite{moffattoth}] :
\begin{equation}
\sigma_\mathrm{LOS}^2(R)=\frac{\int_R^\infty r\sigma^2(r)\rho(r)/\sqrt{r^2-R^2}~dr}{\int_R^\infty r\rho(r)/\sqrt{r^2-R^2}~dr},
\label{eqn17}
\end{equation}
where $R$ is the projected distance between the GC center and the stars being observed.
%%%%%%%%%%%%%%%%%%%%%%%%%%%%%%%%%%%%%%%%%%%%%%%%%%%%%%%%%%%%%%%%%%%%%%%%%%%%%%%%%%%%%%%%%%%%%%%%
%%%%%%%%%%%%%%%%%%%%%%%%%%%%%%%%%%%%%%%%%%%%%%%%%%%%%%%%%%%%%%%%%%%%%%%%%%%%%%%%%%%%%%%%%%%%%%%%
\subsection{Results}
\label{sec2c}
We now fit the observed velocity dispersion profiles for these clusters in the context of Weyl gravity. In our model, the only free parameter remains to be the mass-to-light ratio ($\frac{M}{L}$) which has been assumed to be fixed throughout a given cluster. The best-fit value of the mass-to-light ratio is obtained when the reduced chi-square value is minimized. The reduced chi-square, $\chi_{\nu}^{2}$,  is defined as
\begin{equation}
\chi_{\nu}^{2} =  \frac{1}{f}\sum_{N} \frac{(\sigma_{obs,i} -\sigma_{weyl,i}(\frac{M}{L}))^2}{s_{i}},
\label{eqn18}
\end{equation}
where $f$ is the degrees of freedom, $N$ is the number of data points, $\sigma_{obs,i}$ is the observed velocity dispersion, $\sigma_{weyl,i}(\frac{M}{L})$ is the predicted velocity dispersion given a mass-to-light ratio and $s_{i}$ is the uncertainties in observed velocity dispersion. The best-fit values of mass-to-light ratio for the clusters have been shown in Table \ref{tab2}.\\

We begin with the individual fits of two GCs showing different asymptotic behaviours in the observed dispersion profiles. The first GC is NGC 104, otherwise known as 47 Tuc. For NGC 104, the dispersion data is available up to a distance of 32 pc from the GC center. Observed projected dispersion is found to decline steadily and reaches a value of 3.7 $\pm$ 0.28 km/s. No sign of flattening is observed in data. We approximate the mass model of NGC 104 with a Hernquist profile having total luminosity $L=4.401 \times 10^{5}$ $L_{\odot}$ and half-light radius $r_h=3.57$ pc. We obtain good fit to the data with  $\frac{M}{L}$ = 1.49 (in solar unit) and $\chi_{\nu}^{2}$=0.84 (Figure \ref{fig1}). The best-fit profile for Weyl gravity (red slod line) also shows a continuous fall-off. Interesting point to note here is that the Weyl gravity fit and the Newtonian contribution has little difference in the interior (r$<$10 pc) and deviates slightly beyond that.\\

\begin{figure}
	\centering
	\includegraphics[width=1.0\linewidth]{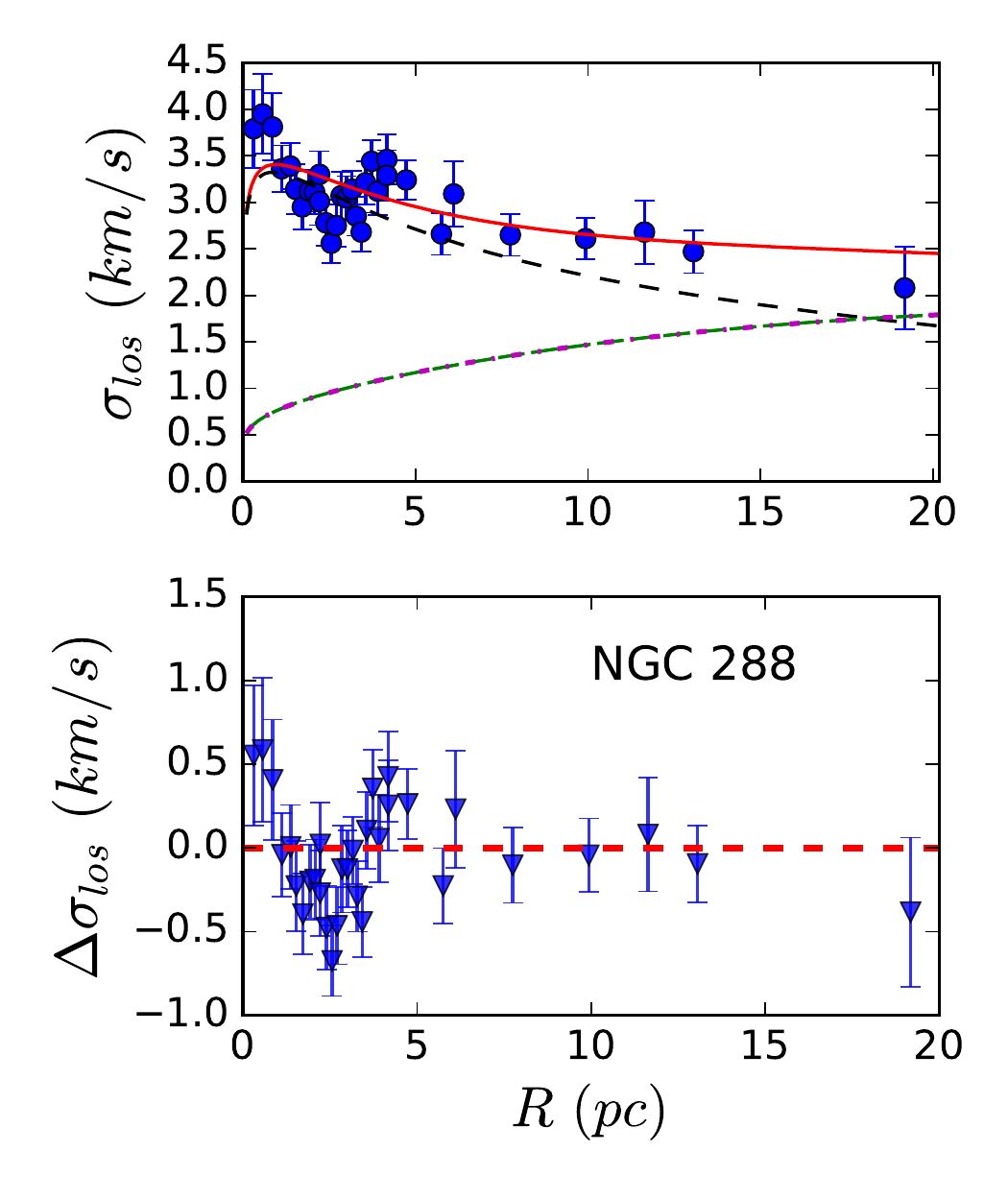}
	\caption[]{\textbf{\textit{(Upper panel)} Velocity dispersion profile of the  NGC 288 have been plotted as a function of projected radial distances. Additionally, we show the contribution due to the Newtonian term alone in black dashed line and the contribution from the two linear terms only in long dashed green and the contribution from the linear and quadratic terms together in dotted magenta.\textit{(Lower panel)} The figure shows the residual of fit as a function of radial distances from the cluster center.}}
	\label{fig2}
\end{figure}

The second GC we study is NGC 288. It is a low concentration cluster and is located at a distance of 9.8 kpc. For this particular cluster, dispersion data is available up to 18 pc from the cluster center. The cluster shows a recognizable trend of flattening of the dispersion profile in the outer region. The dispersion profile settles for an asymptotic value of  2.5 $\pm$ 0.3 km/s beyond 6 pc from the cluster center. The best fit Weyl gravity model is obtained for $\frac{M}{L}$ = 1.94 (in solar unit) with $\chi_{\nu}^{2}$=0.35 (Figure \ref{fig2}). Unlike in NGC 104, we find a significant difference between the Weyl gravity fit and the Newtonian contribution in all radial distances for this particular GC.\\

\begin{figure*}
	\centering
	\subfigure[]{\label{fig3a}
		\includegraphics[scale=0.562]{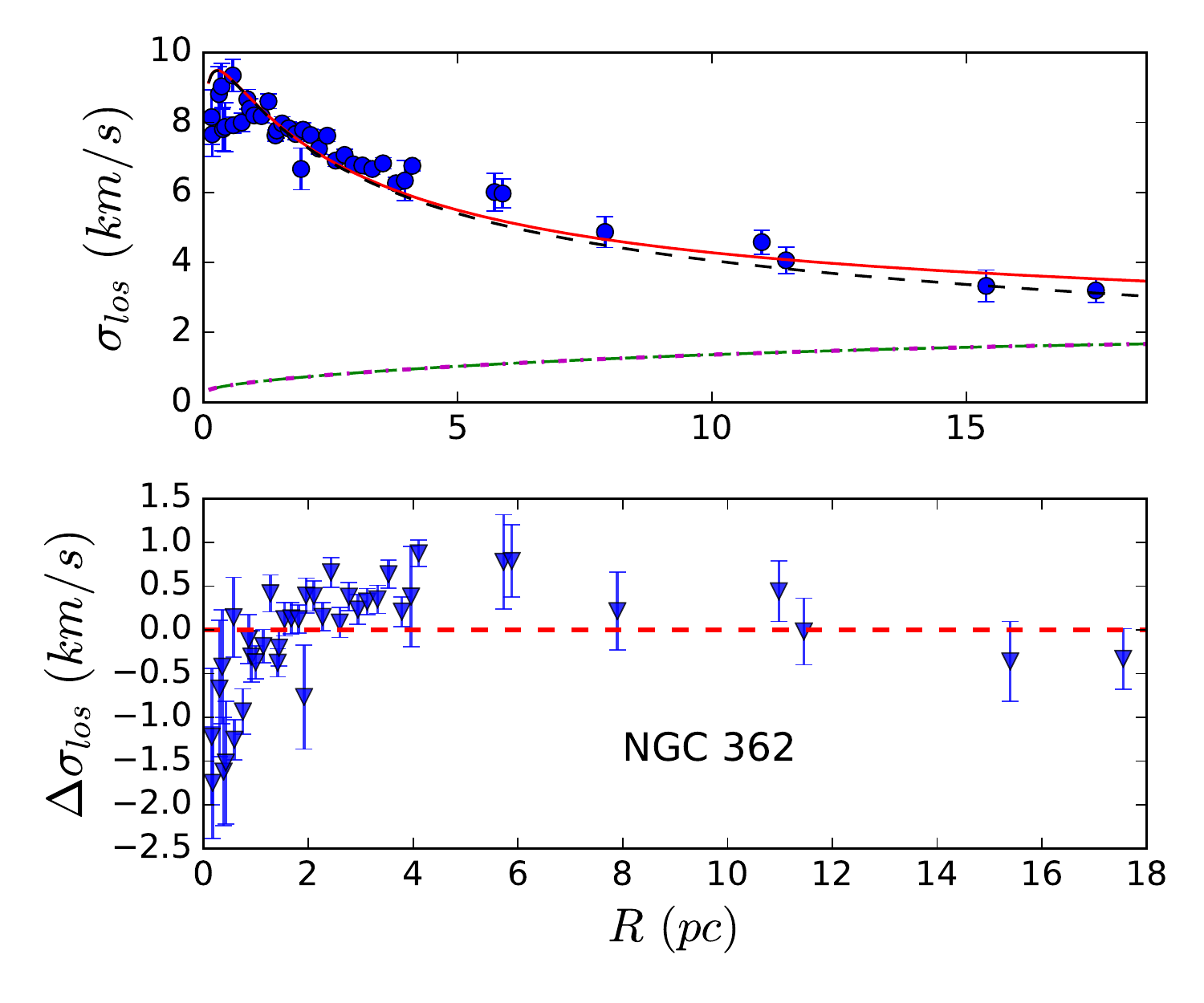}}
	\subfigure[]{\label{fig3b}
		\includegraphics[scale=0.562]{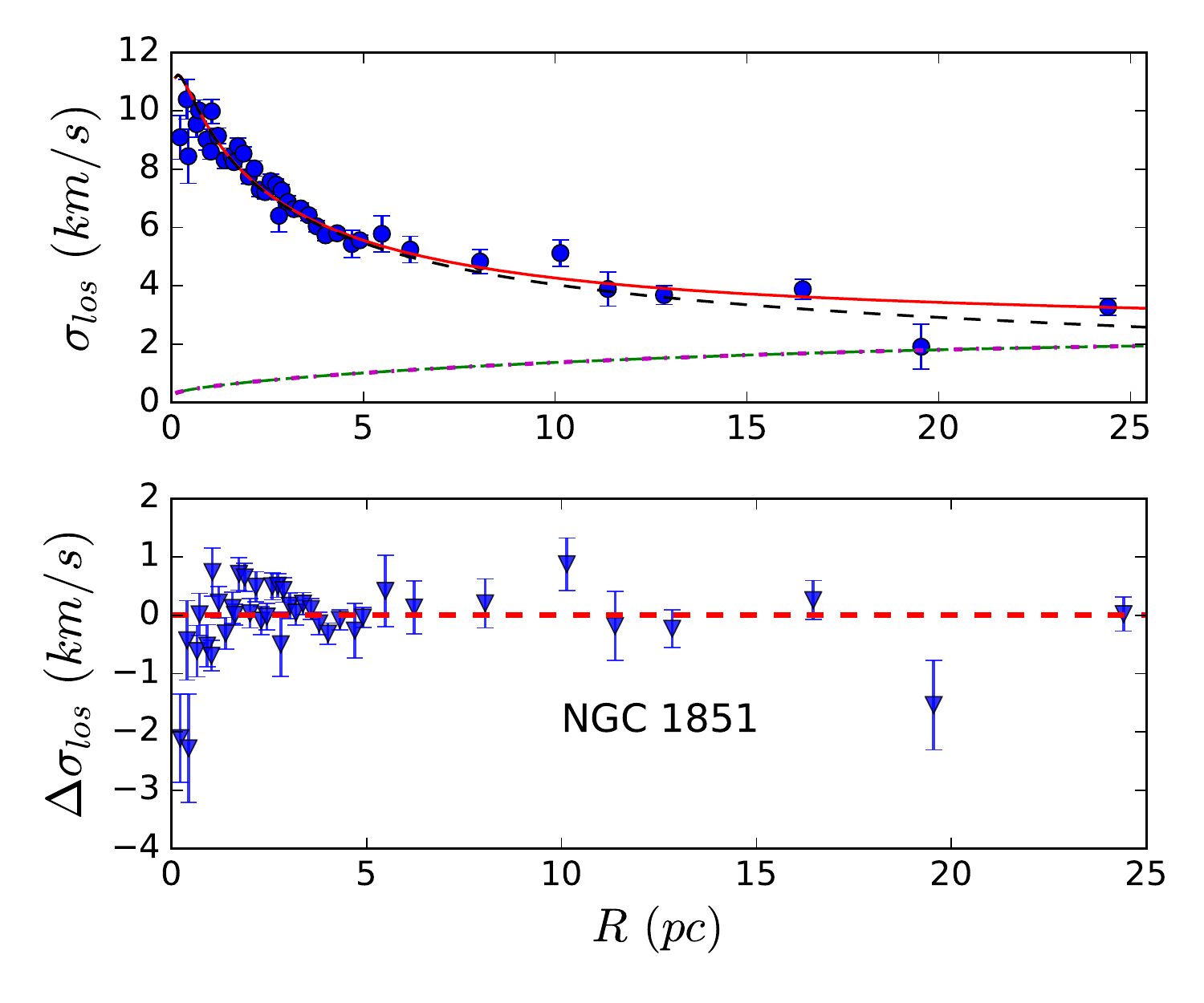}}\\
	\subfigure[]{\label{fig3c}
		\includegraphics[scale=0.562]{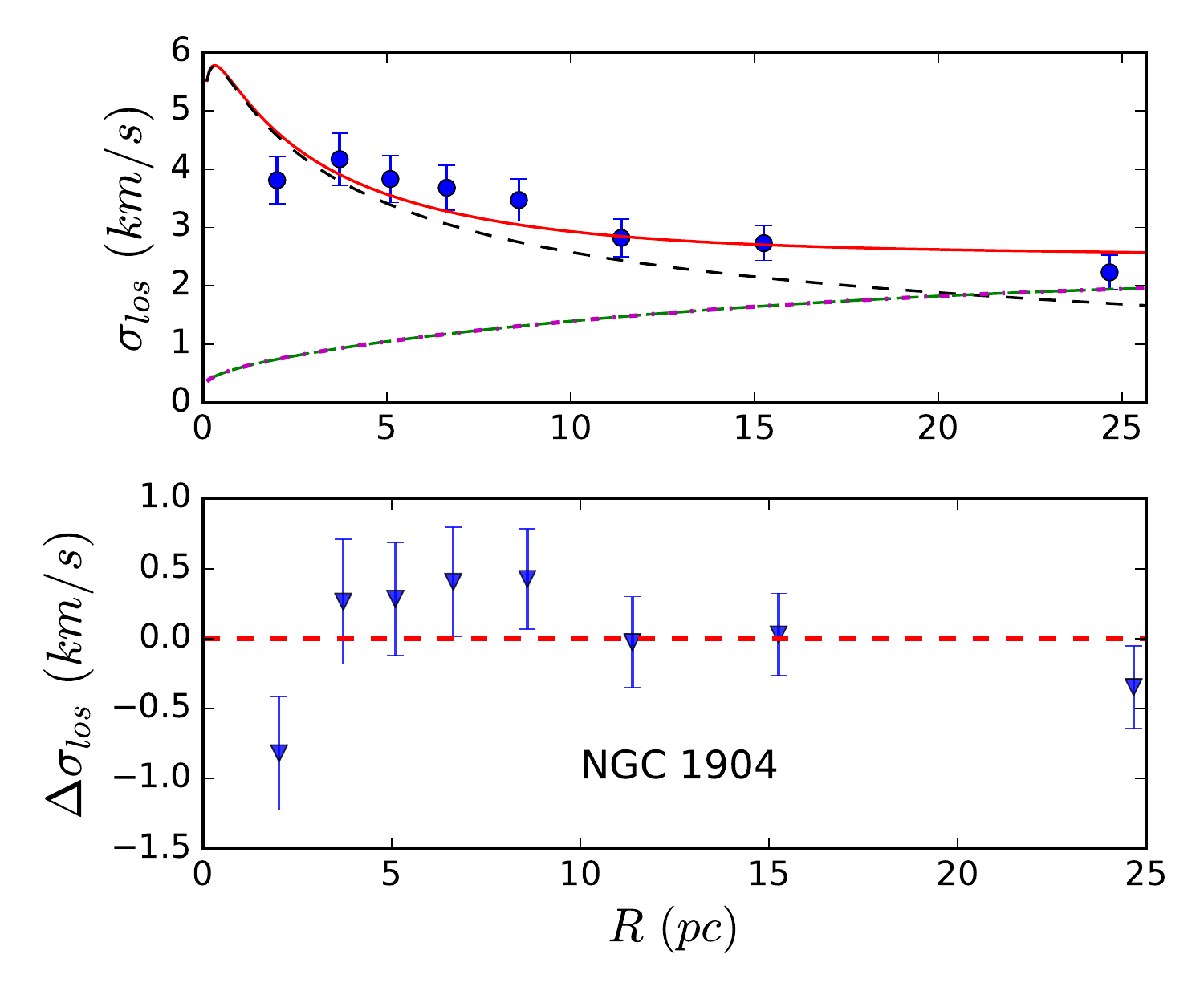}}
	\subfigure[]{\label{fig3d}
		\includegraphics[scale=0.562]{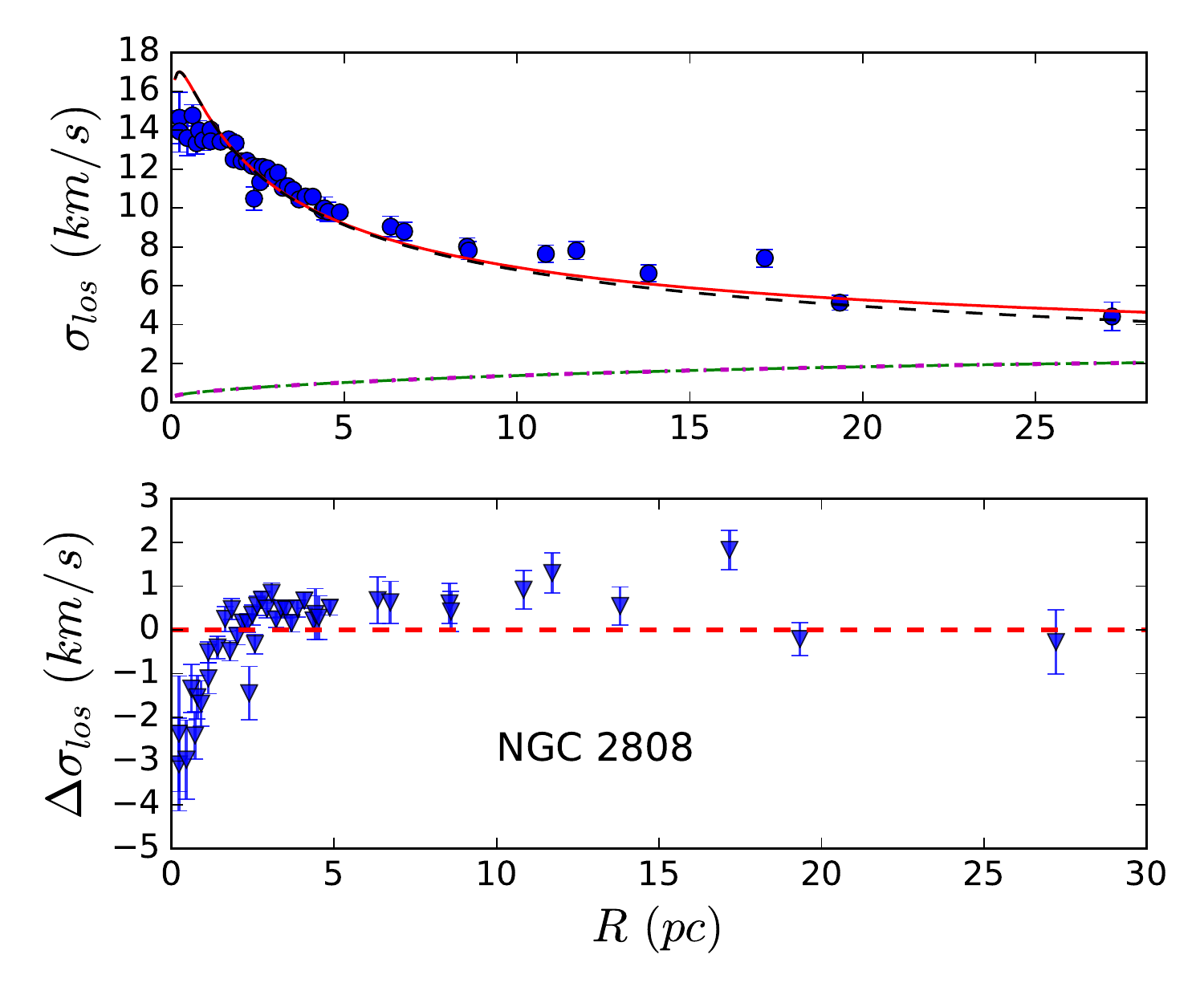}}\\
	\subfigure[]{\label{fig3e}
		\includegraphics[scale=0.562]{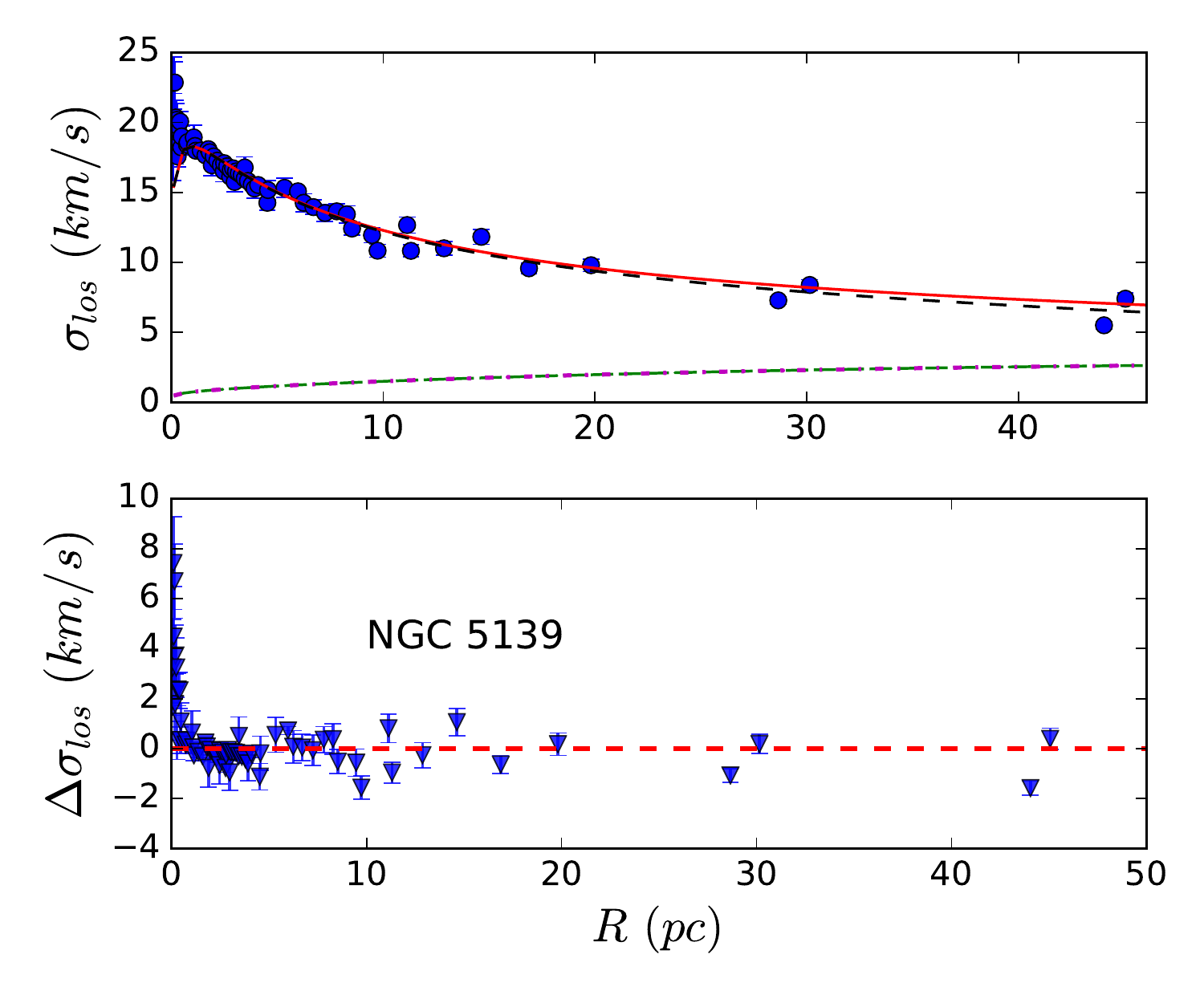}}
		\subfigure[]{\label{fig3f}
			\includegraphics[scale=0.562]{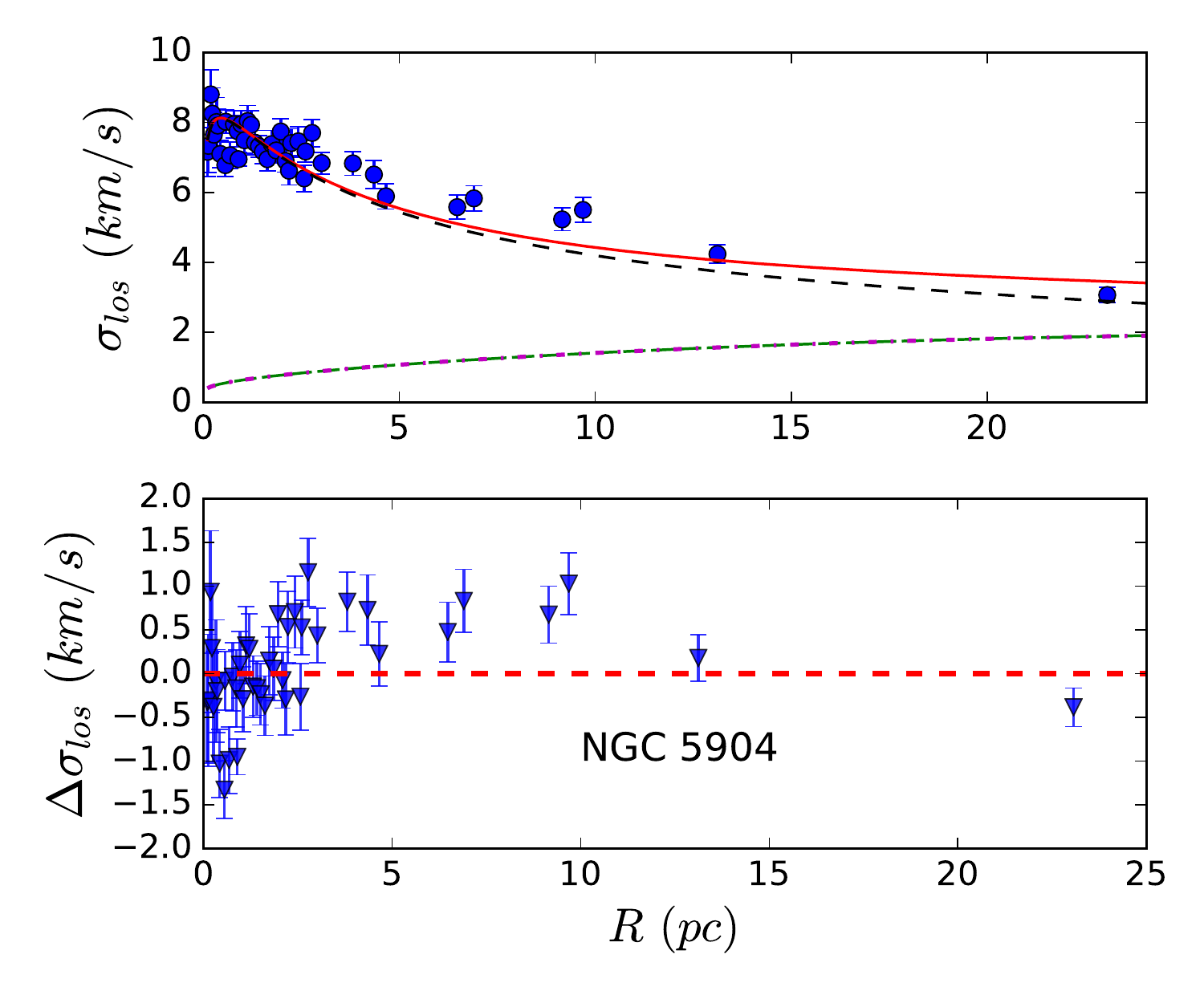}}
	\caption{\label{fig3} Same as Fig. \ref{fig2} but for : (a) NGC 362, (b) NGC 1851, (c) NGC 1904, (b) NGC 2808, (c) NGC 5139 \& (d) NGC 5904. }
\end{figure*}

\begin{figure*}
	\centering
	\subfigure[]{\label{fig4a}
		\includegraphics[scale=0.562]{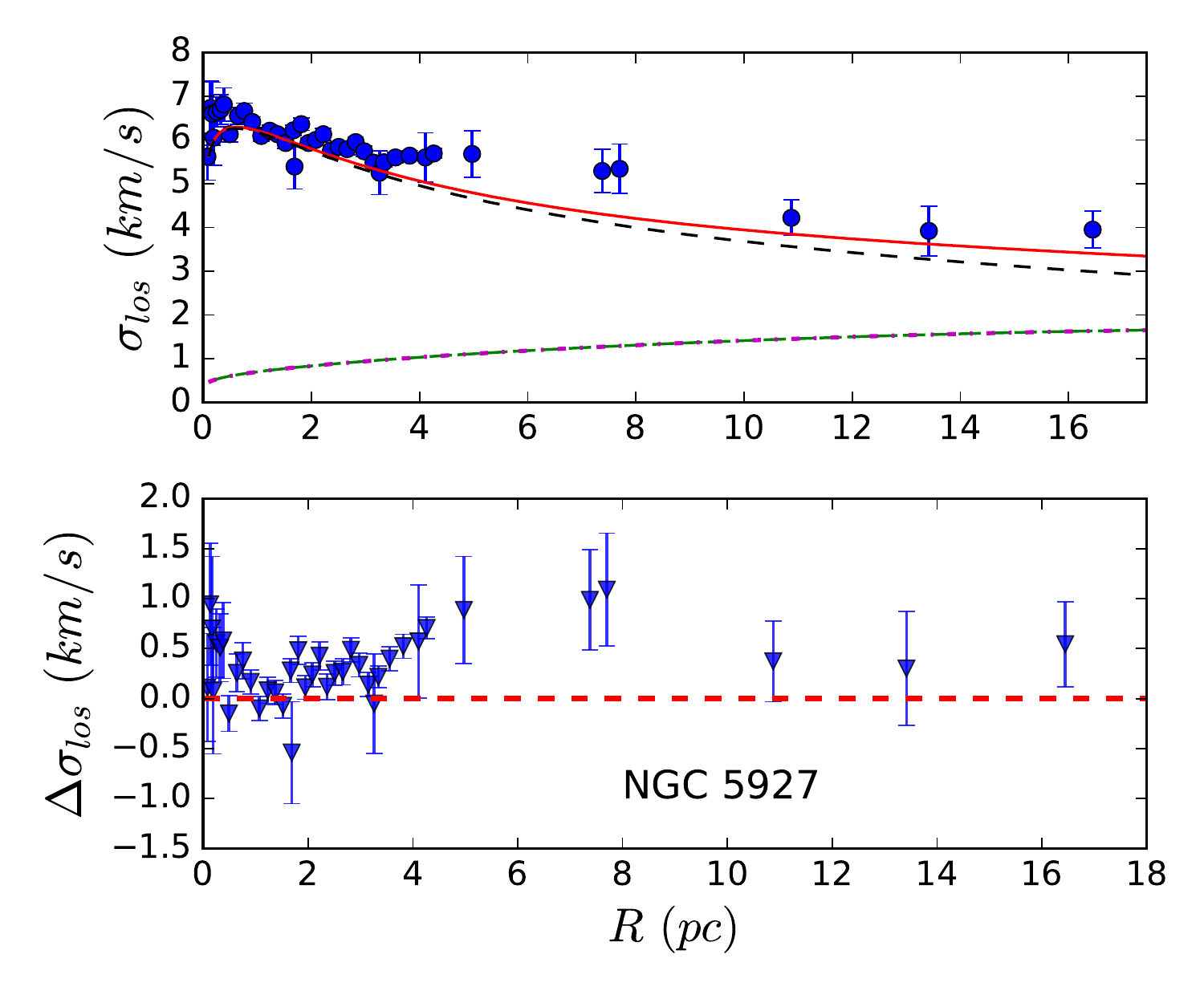}}
	\subfigure[]{\label{fig4b}
		\includegraphics[scale=0.562]{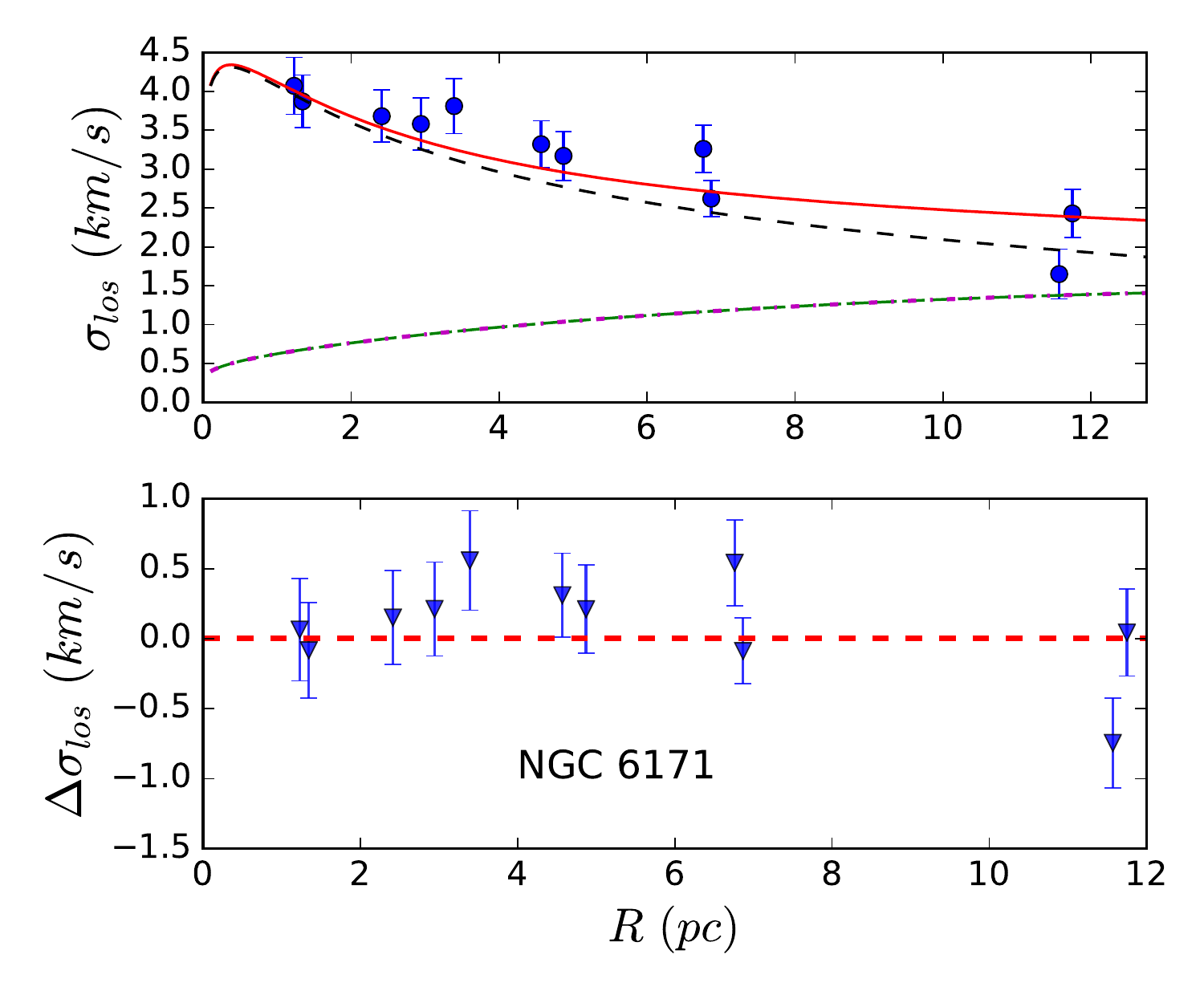}}\\
	\subfigure[]{\label{fig4c}
		\includegraphics[scale=0.562]{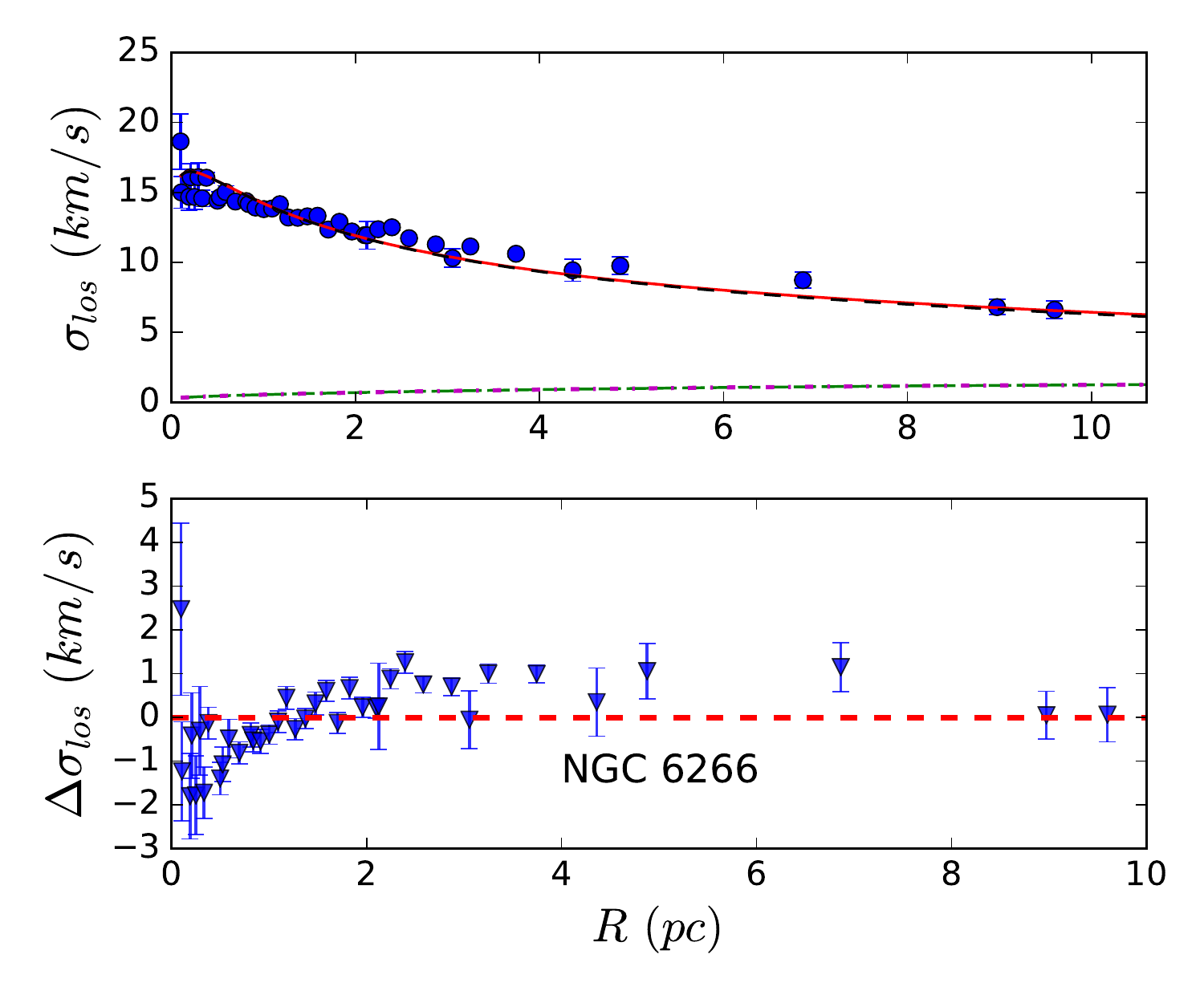}}
	\subfigure[]{\label{fig4d}
		\includegraphics[scale=0.562]{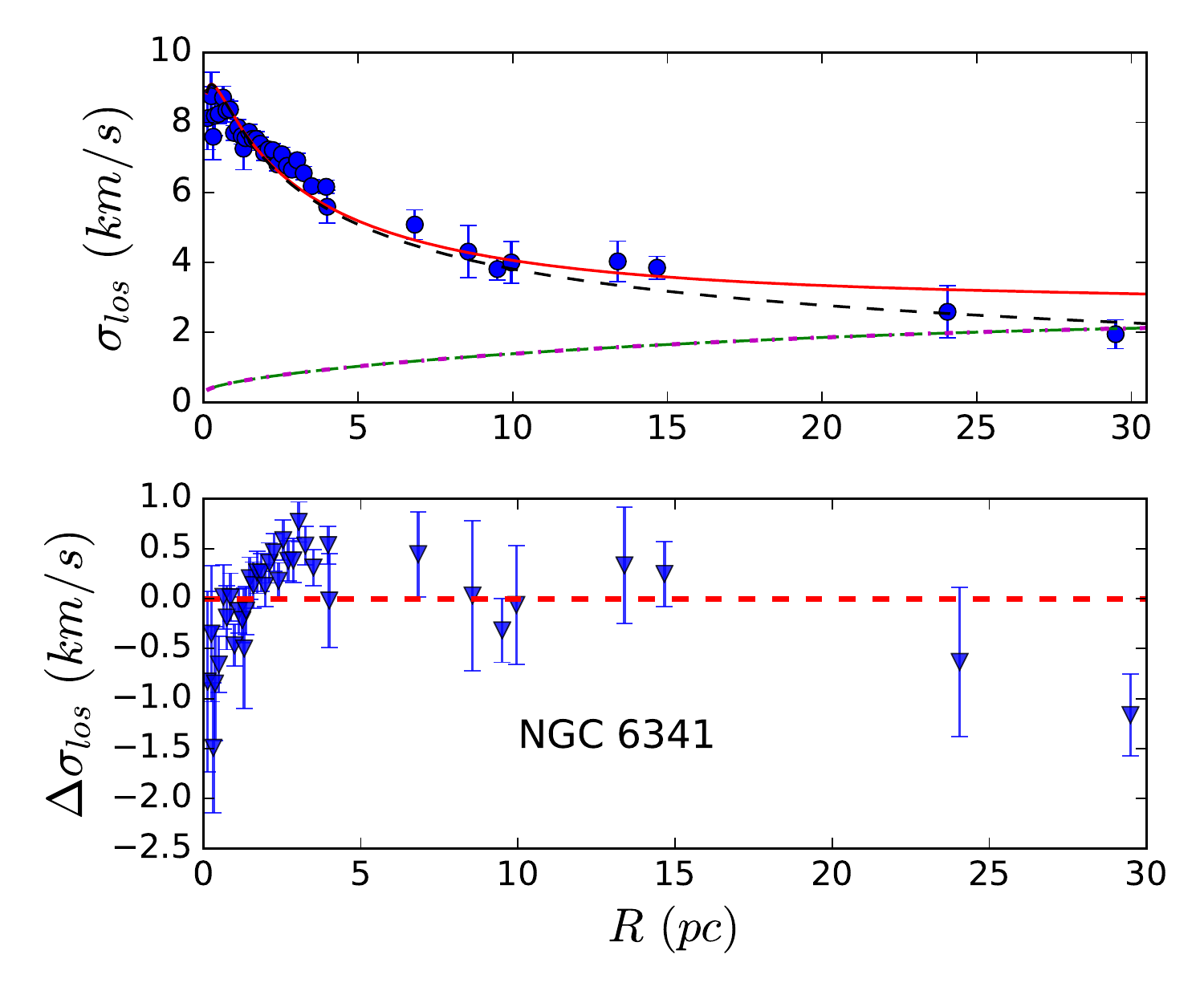}}\\
	\subfigure[]{\label{fig4e}
		\includegraphics[scale=0.562]{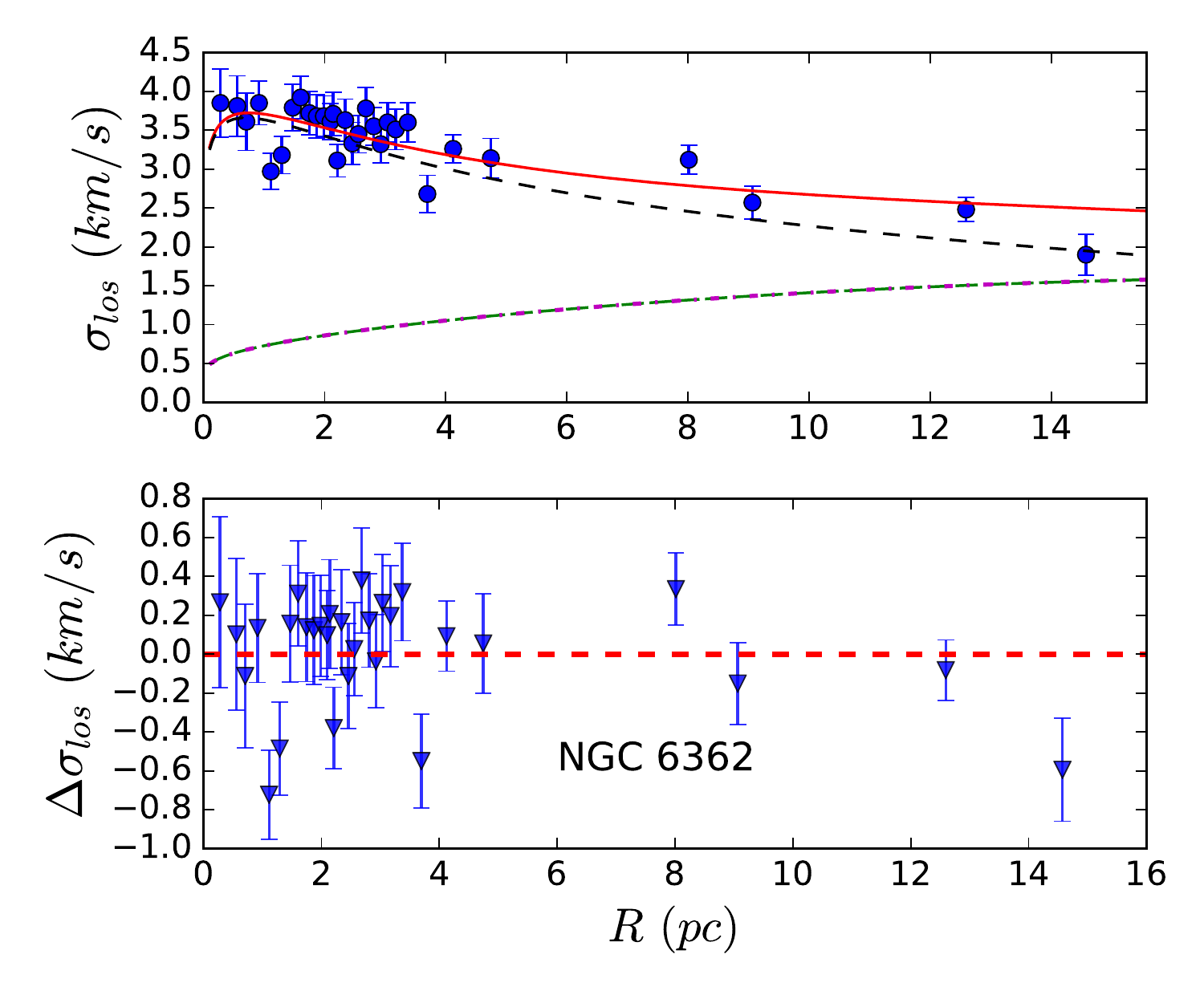}}
	\subfigure[]{\label{fig4f}
		\includegraphics[scale=0.562]{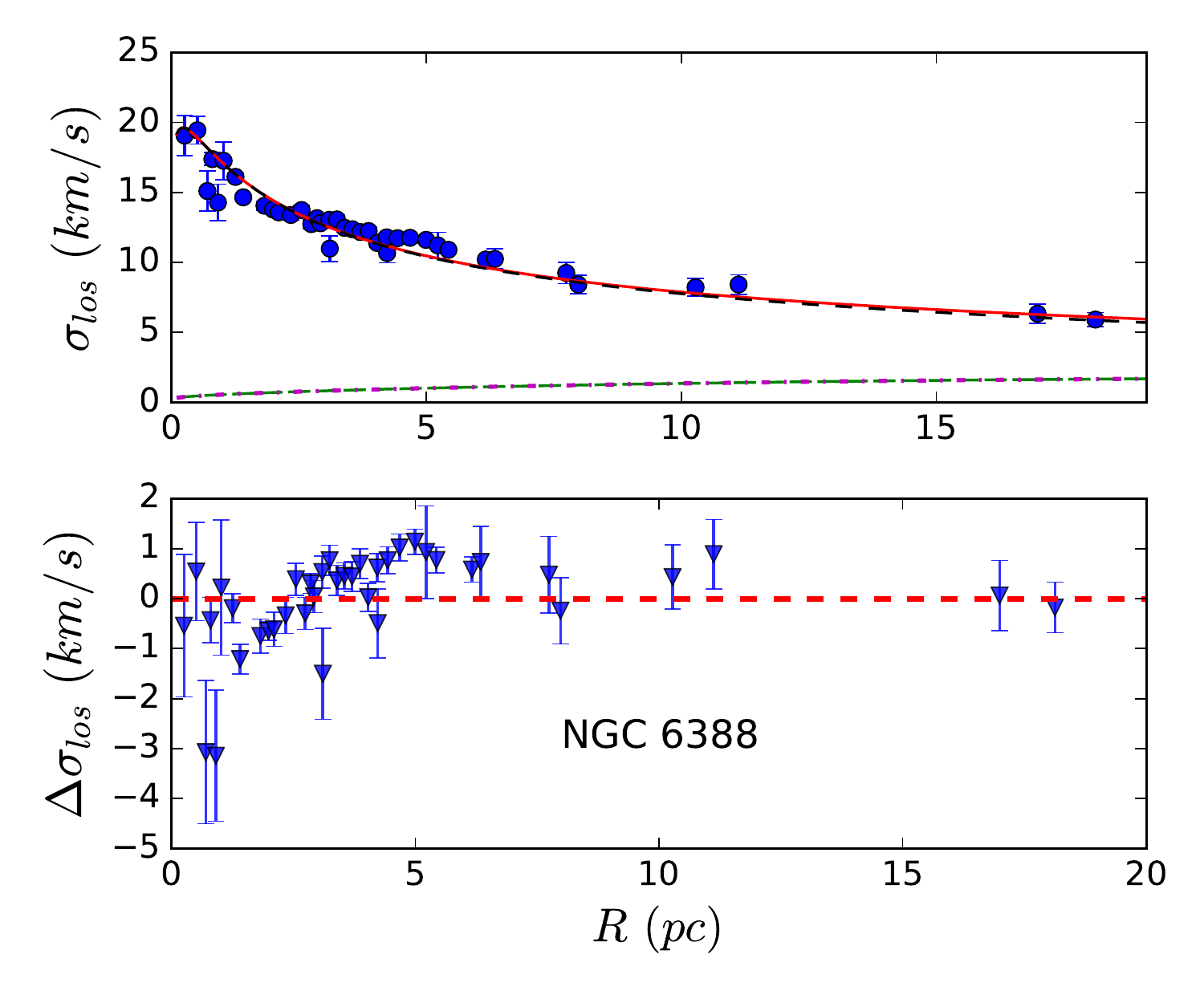}}
	\caption{\label{fig4} Same as Fig. \ref{fig2} but for : (a) NGC 5927, (b) NGC 6171, (c) NGC 6266, (b) NGC 6341, (c) NGC 6362 \& (d) NGC 6388. }
\end{figure*}

\begin{figure*}
	\centering
	\subfigure[]{\label{fig5a}
		\includegraphics[scale=0.562]{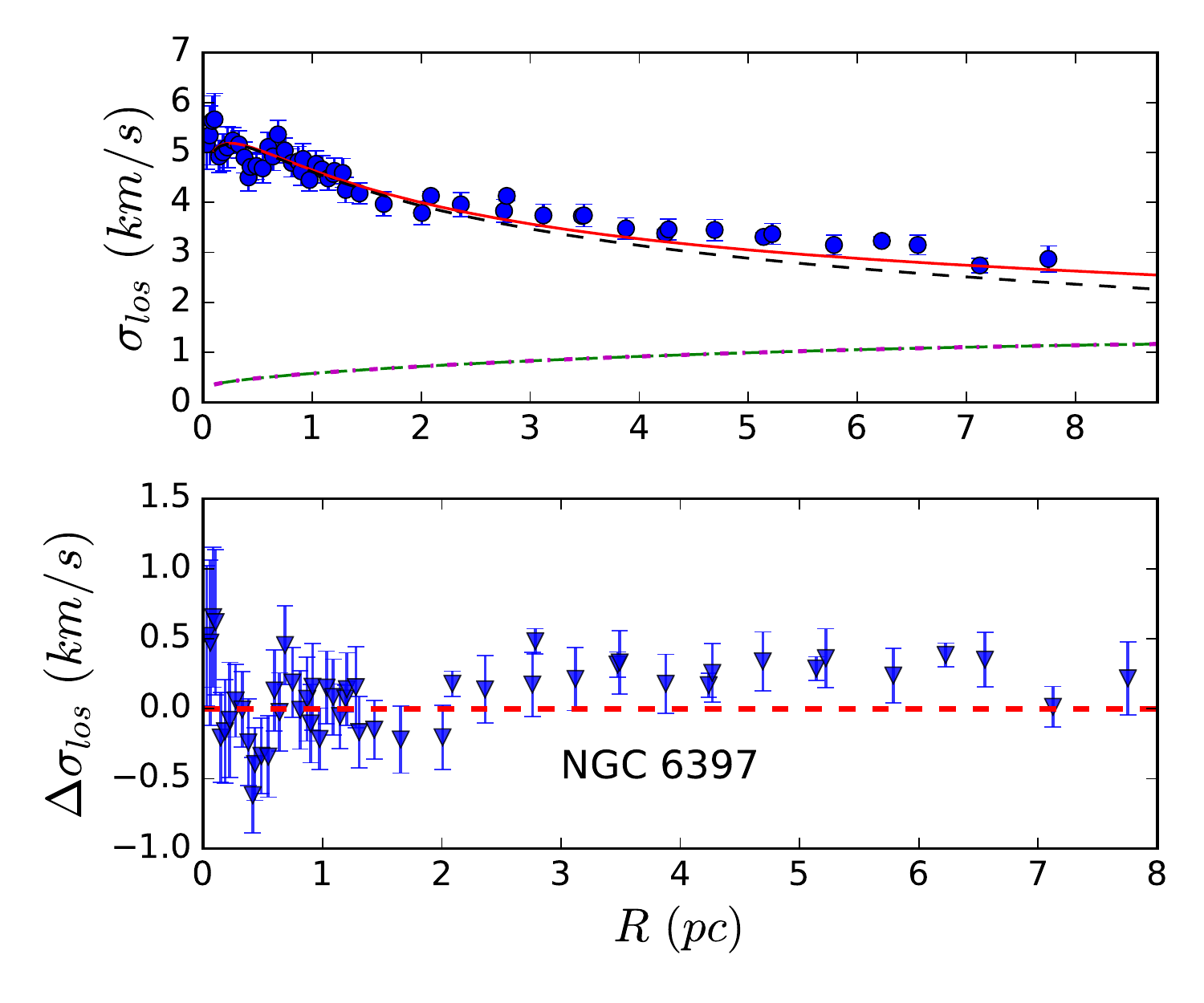}}
	\subfigure[]{\label{fig5b}
		\includegraphics[scale=0.562]{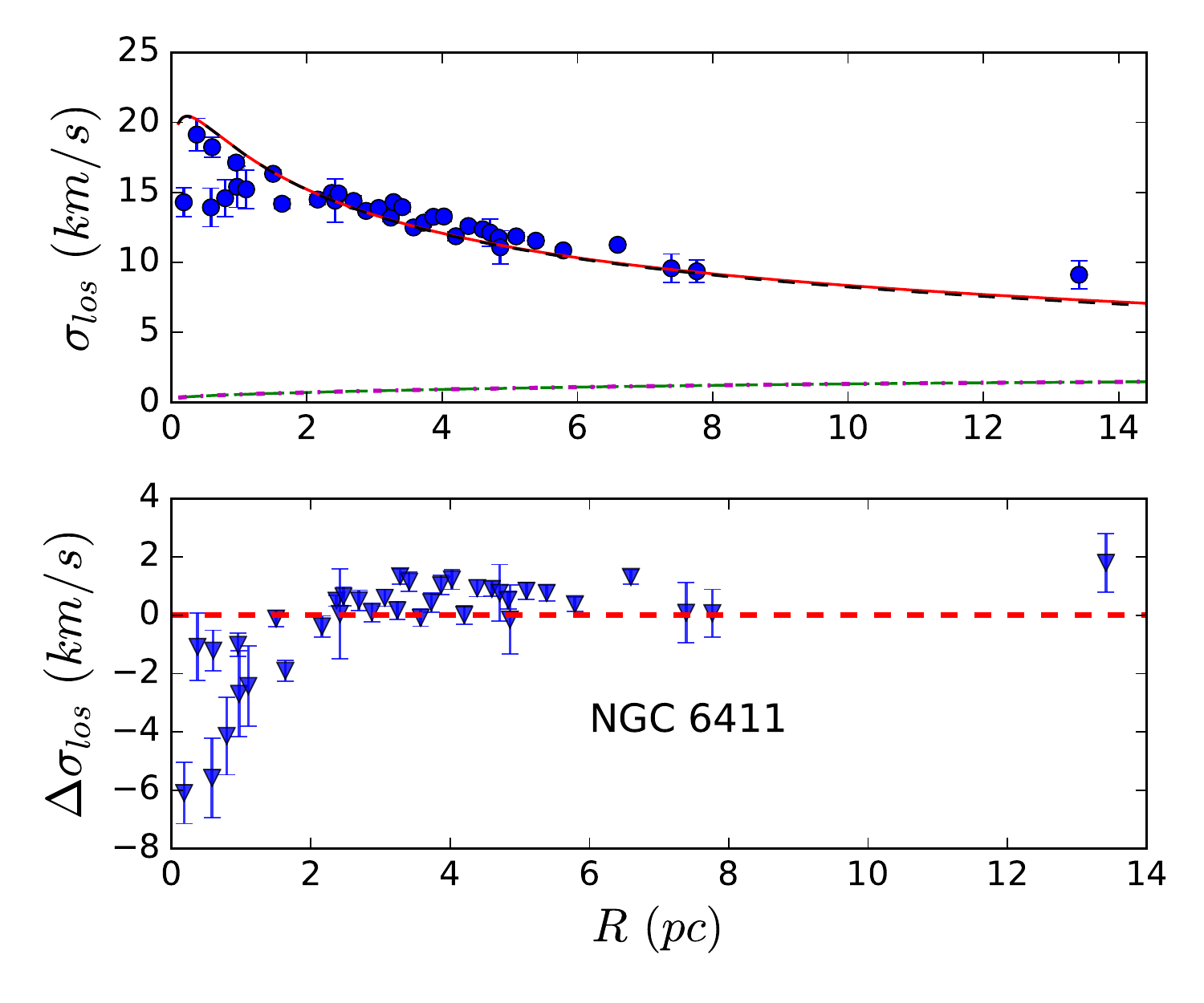}}\vskip0.5mm
	\subfigure[]{\label{fig5c}
		\includegraphics[scale=0.562]{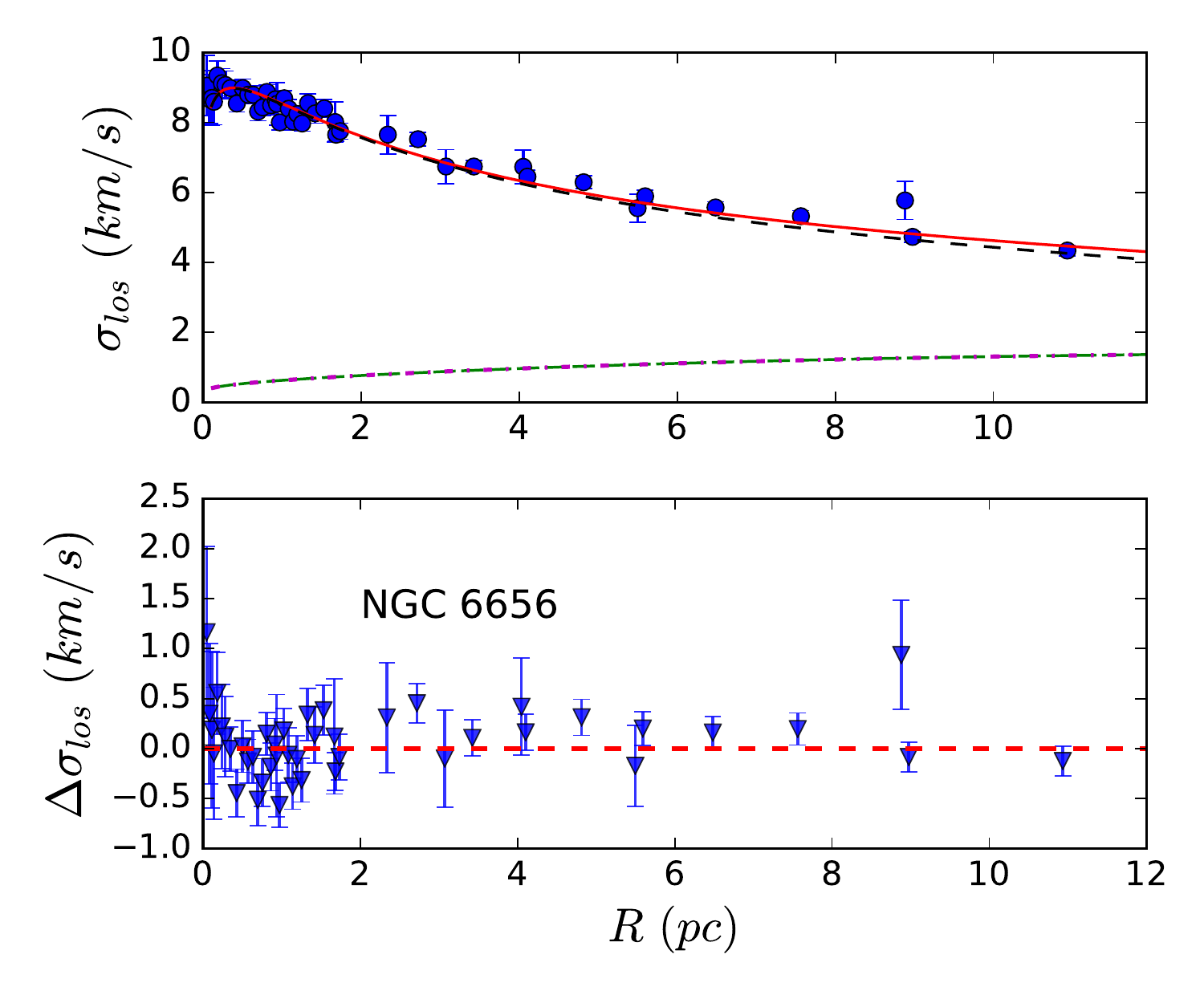}}
	\subfigure[]{\label{fig5d}
		\includegraphics[scale=0.562]{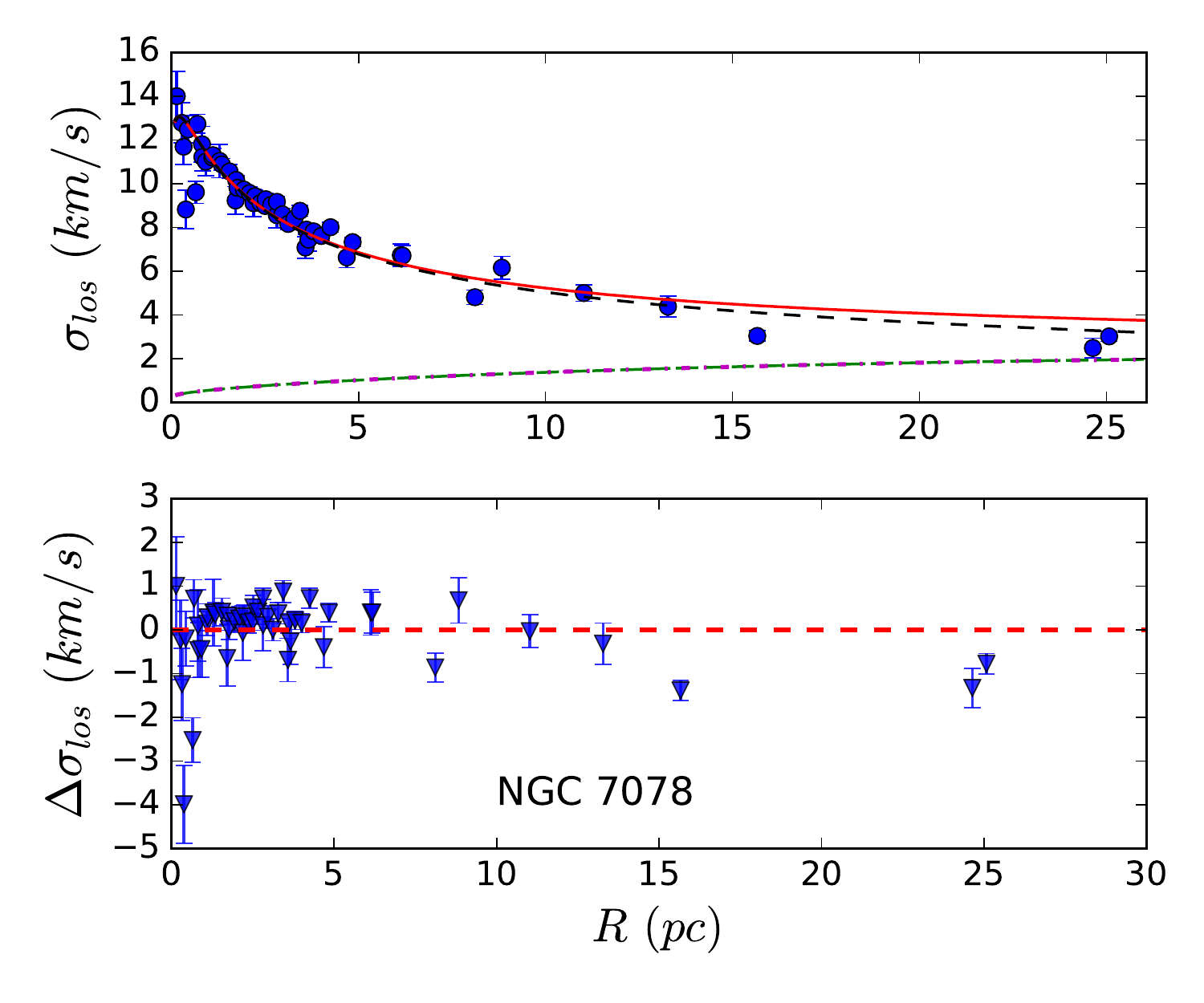}}\\
	\subfigure[]{\label{fig5e}
		\includegraphics[scale=0.562]{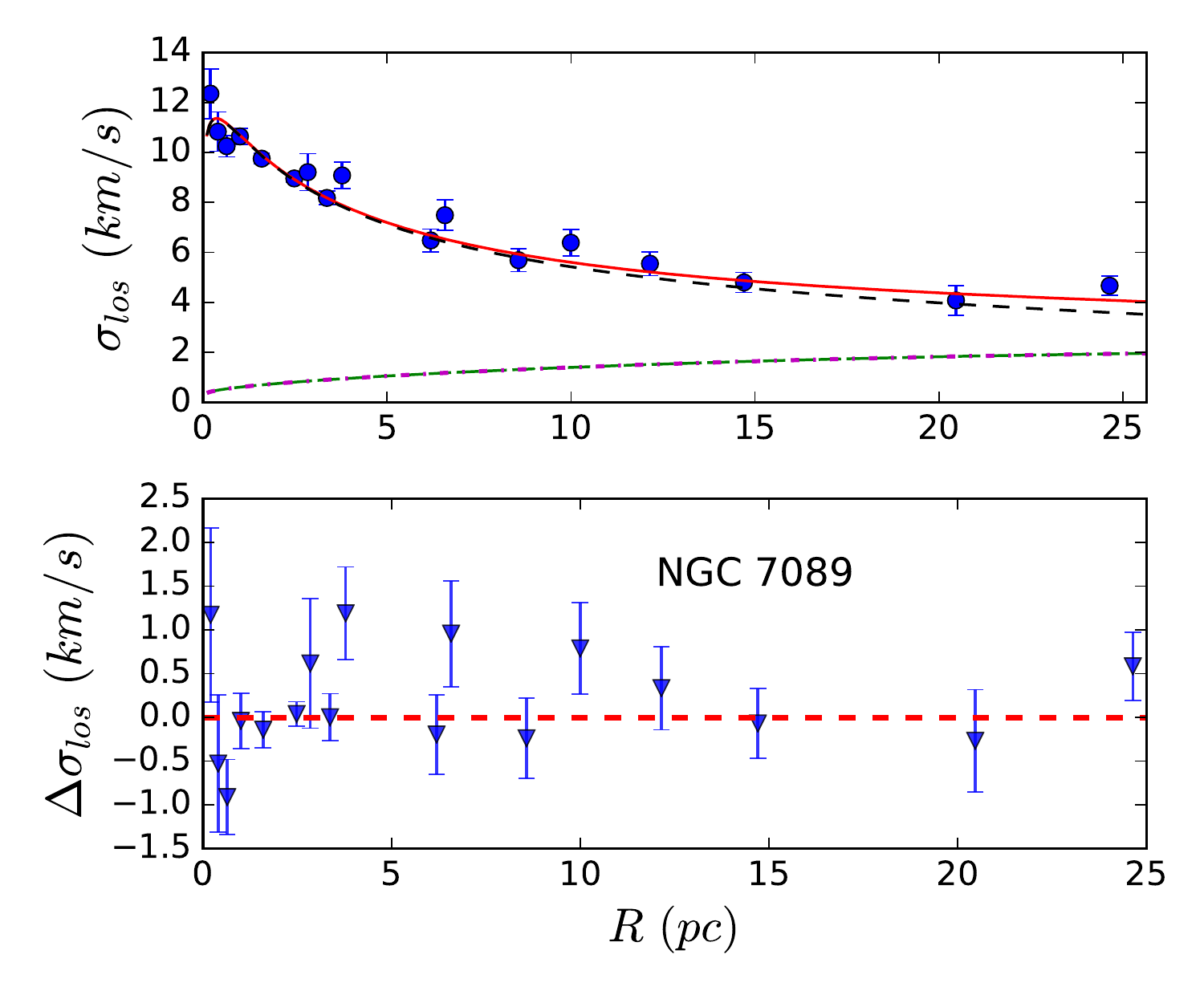}}
	\subfigure[]{\label{fig5f}
		\includegraphics[scale=0.562]{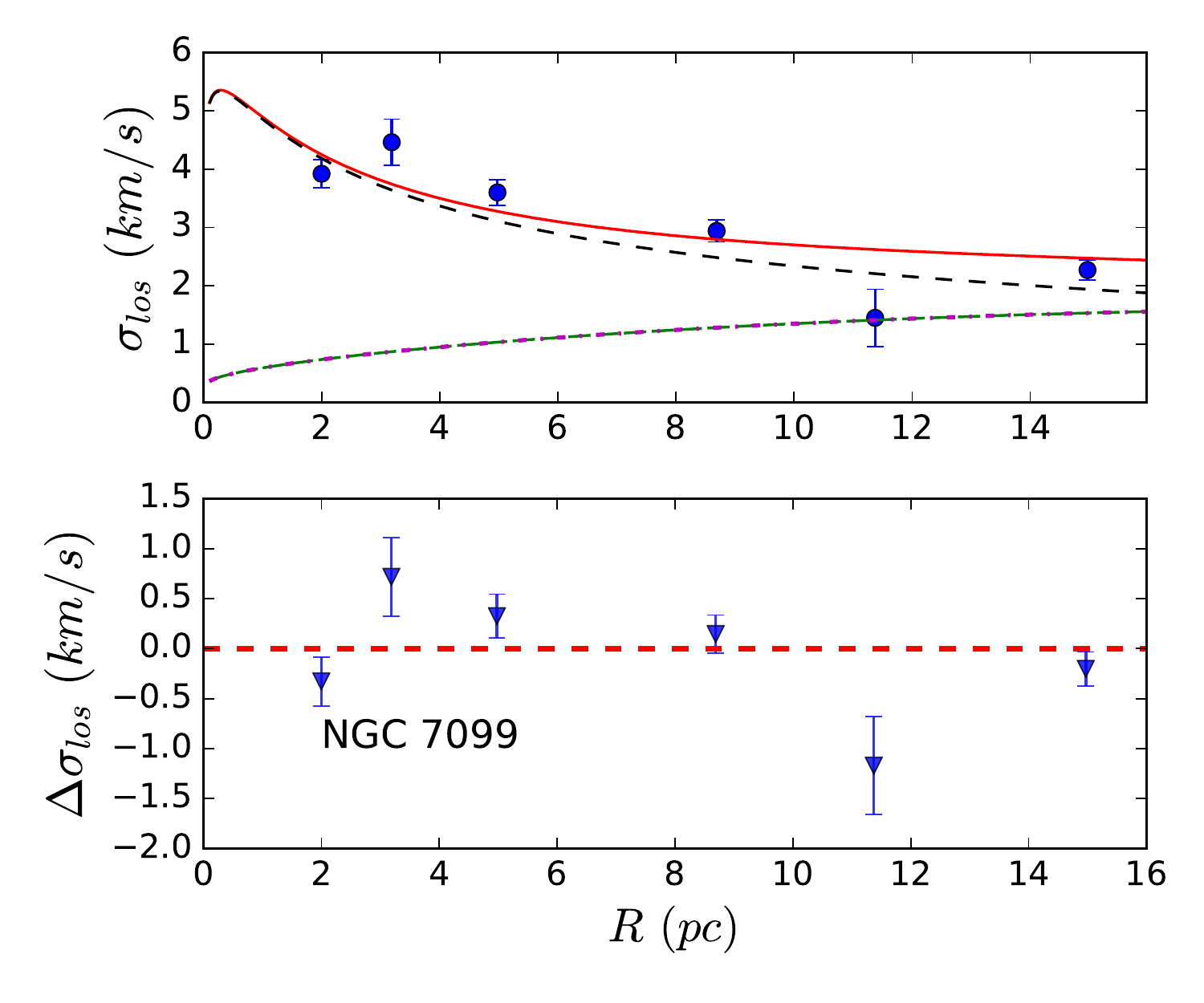}}
	\caption{\label{fig5} Same as Fig. \ref{fig2} but for : (a) NGC 6397, (b) NGC 6411, (c) NGC 6656, (b) NGC 7078, (c) NGC 7089 \& (d) NGC 7099. }
\end{figure*}

The fits to the velocity dispersions of other 18 GCs are shown in Figure  \ref{fig3}, \ref{fig4} and \ref{fig5}. Most of the best-fit profiles yield a  $\chi_{\nu}^{2} \sim 1.00$ indicating good agreements between data and model. For some GCs (e.g. NGC 2808, NGC 6266, NGC 6411), the  $\chi_{\nu}^{2}$ values of the best fit are relatively larger than those obtained from the other clusters but are still acceptable. We also find that the contribution from the quadratic term (fourth term in Eq. \ref{eqn8})  is not significant at the scale of globular clusters as there is hardly any difference between the contribution from the two linear terms alone (long dashed green) and the linear and quadratic terms combined (dotted magenta).

%5++++++++++++++++++++++++++
\begin{table}
	\caption{Weyl gravity fits : The first three entries give the name of the GC, best-fit value for the mass-to-light ratio and best fit reduced chi-square value. The fourth column gives the inferred total mass of the GC obtained via Weyl gravity fit. The fifth column gives the total mass inferred from stellar population synthesis \citep{mclaughlin2005resolved}.}
	\label{tab2}
	\begin{center}
		\begin{tabular}{lcccc}
			\hline
			GC  &best-fit $\frac{M}{L}$ &$\chi_{\nu}^{2}$ & $log_{10}(\frac{M_{Weyl}}{M_{\odot}})$ &$log_{10}(\frac{M_{PS}}{M_{\odot}})$\\
			\hline
			NGC 104 & 1.49 &0.84 &5.81 & 6.3 $\pm$ 0.2\\
			NGC 288 & 1.94 & 0.35 & 4.97 &4.8 $\pm$ 0.2\\
			NGC 362 & 1.20&1.25 &5.39 &\\
			NGC 1851 & 1.60 &0.82 &5.37 &5.6 $\pm$ 0.2\\
			NGC 1904 & 1.04 &0.47 & 5.01& 4.9 $\pm$ 0.2\\
			NGC 2808 & 1.53 & 2.20 &5.83 & 5.9 $\pm$ 0.2\\
	    	NGC 5139 & 2.32 &2.63 &6.4 &6.4 $\pm$ 0.2\\
			NGC 5904 & 1.17 &0.94 &5.45 & 5.6 $\pm$ 0.2\\
			NGC 5927 & 1.75 &0.84 &5.37 & \\
			NGC 6171 & 1.75 &0.44 &4.84 & 4.9 $\pm$ 0.2\\
			NGC 6266 & 2.20 &1.72 &5.78 &\\
			NGC 6341 & 1.48 &0.78 &5.34 &5.3 $\pm$ 0.2\\
			NGC 6362 & 1.67 & 0.35 &4.97 &\\
			NGC 6388 & 1.64 &1.49 &5.95 & 6.1 $\pm$ 0.2\\
			NGC 6397 & 1.75 &0.36 &4.85 & \\
			NGC 6411 & 1.71 & 4.03 &6.01 & 6.2 $\pm$ 0.2\\
			NGC 6656 & 1.66 &0.33 &5.50 &5.4 $\pm$ 0.2\\
			NGC 7078 & 1.11&1.49 &5.57 &\\
			NGC 7089 & 1.29 &0.68 &5.66 &5.5 $\pm$ 0.2\\
			NGC 7099 & 1.17 &1.11 &4.92 &4.8 $\pm$ 0.2\\
			\hline
		\end{tabular}
	\end{center}
\end{table}
%5++++++++++++++++++++++++++

\subsubsection{Best-fit mass-to-light ratios}
Our analysis finds good fit to the observed velocity dispersions for 20 GCs with best-fit mass-to-light ratio in the range 1.0 $< \frac{M}{L} <$ 2.4 with $(\frac{M}{L})_{avg}=1.55$, which lies in the lower end of values predicted by stellar evolution models \citep{bressan2012}. However, \cite{kruijssen2009} found that, due to various dynamical effects, such as preferential retention of remnants and progressive depletion of the mass function, the estimated dynamical $\frac{M}{L}$ would be systematically smaller than the ones predicted in stellar population models \citep{baumgardt2018mean}. Our estimation of (dynamical) mass-to-light ratios for the GCs are thus reasonable.\\

\begin{figure}
	\centering
	\includegraphics[width=0.8\linewidth]{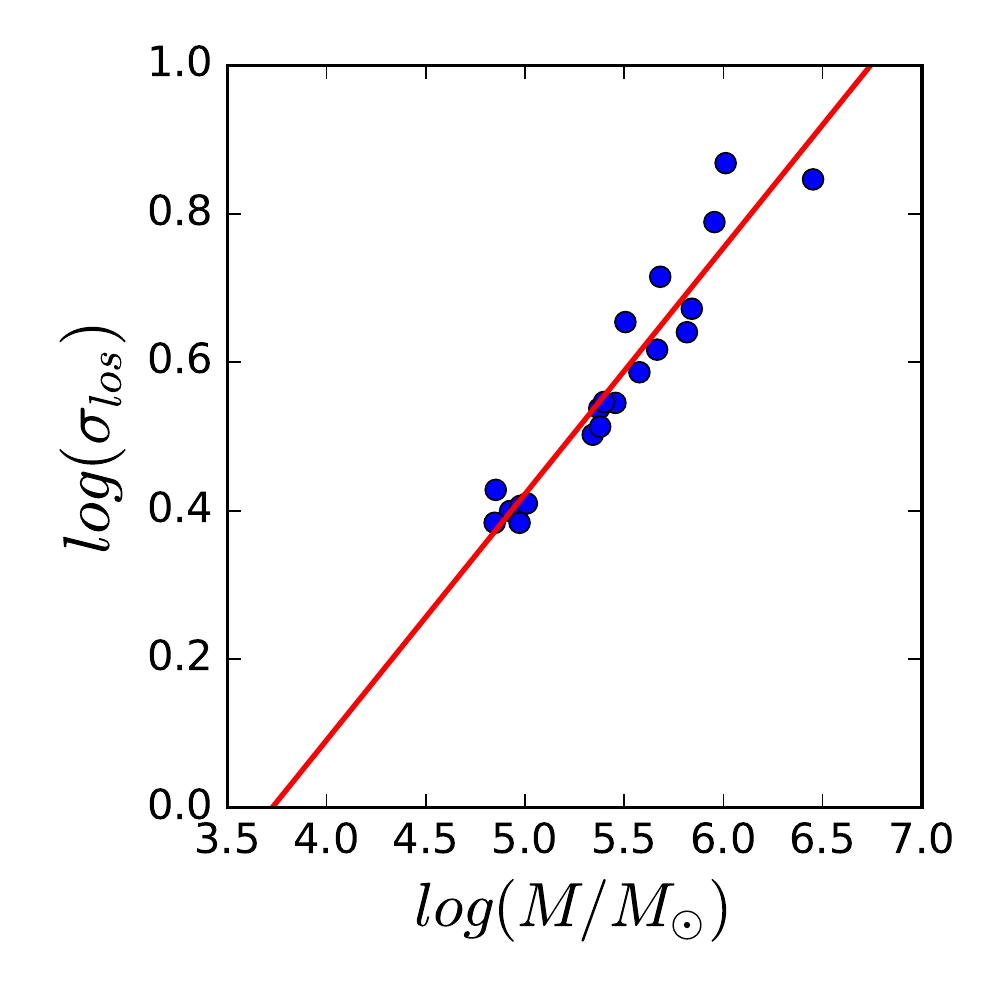}%{gc_mass}
	\caption[]{\textbf{We plot the asymptotic velocity dispersion of the best-fit Weyl gravity profile for each cluster as a function of total mass. The line gives the best fit $\sigma$ $\propto$ $M^{q}$ scaling for the data. $q$ is found to be 0.32 $\pm$ 0.03, which is similar to the galactic Tully-Fisher relation.} }
	\label{fig6}
\end{figure}

\subsubsection{Nature of the dispersion profiles and Weyl gravity}
We note that the velocity dispersion profiles of 20 GCs studied here show three different kinds of characteristics. Most of the GCs exhibit a monotonous decline in dispersion while NGC 1851, NGC 6341, NGC 7089 and NGC 7099 show mild degree of flattening in the outer region of the cluster. For NGC 288 and NGC 1904, the flattening is relatively more prominent.  Our study finds that Weyl gravity models can reasonably capture this diversity in dispersion profiles. In case of NGC 6341, the flattening is easily recognizable in the best-fit profile while the data has a slight declining feature. \\

\subsubsection{Tully-Fisher like relation in GCs}
Finally, we plot, in loglog scale, the asymptotic value of velocity dispersion for each cluster as a function of derived total mass for all 20 GCs studied in this paper (Figure \ref{fig6}).  The total mass for a particular cluster has been inferred through chi-square fit to the data within the framework of Weyl gravity. We choose the best-fit Weyl gravity dispersion value corresponding to the last measured data point for each GC to represent the asymptotic velocity dispersion. The best fit solid line has a slope 0.32 $\pm$ 0.03 which is remarkably close to the value of 0.25, predicted in generic modified gravity theories and Tully-Fisher relation at the galactic scale ( $\sigma$ $\propto$ $M^{1/4}$ ). This plot thus strengthens the argument that the flattening of dispersion profiles in some GCs might be a signature of modified gravity and bolsters the case for Weyl gravity in the GCs. 

\subsubsection{Robustness of results}
\label{sec334}
\begin{figure}
	\centering
	\includegraphics[width=0.85\linewidth]{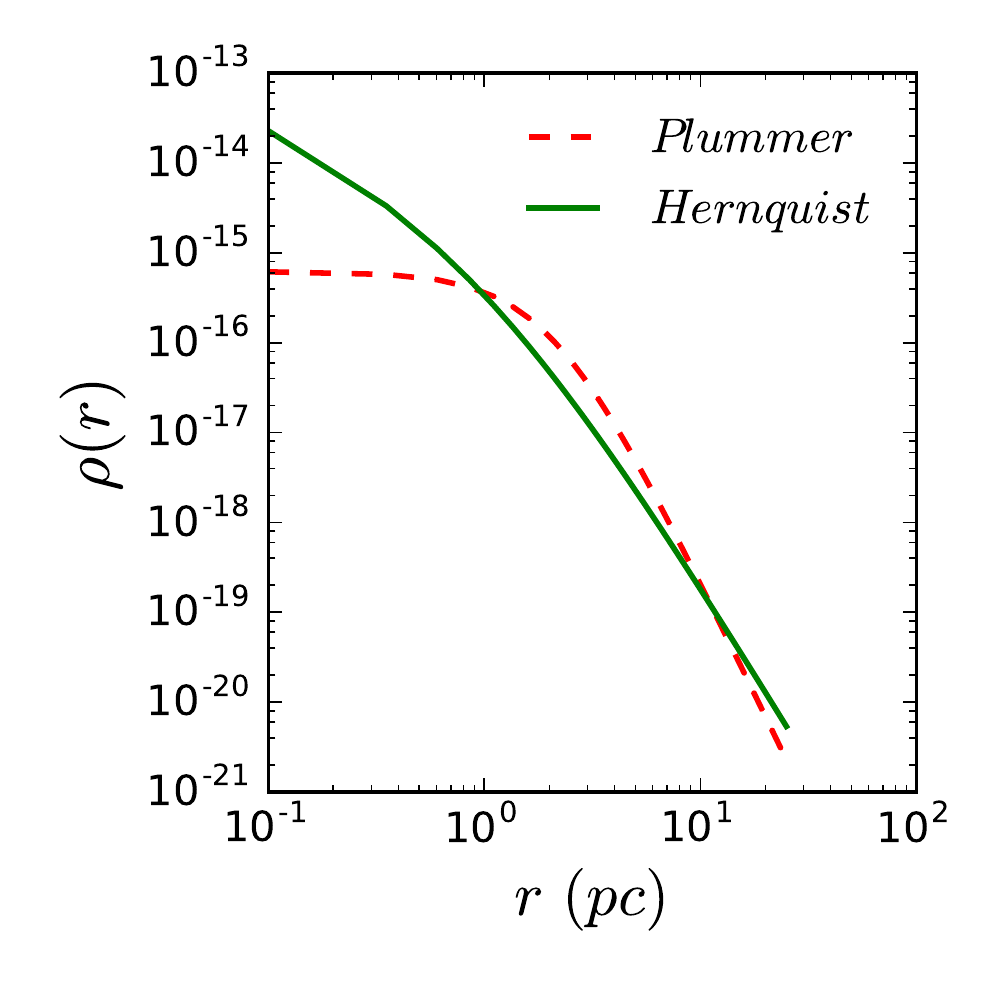}%{gc_mass}
	\caption[]{\textbf{We plot the mass density of NGC 1851 for two different mass models: Hernquist profile (see Section \ref{sec2a}) in green solid line \& Plummer profile (see Section \ref{sec334}) in red dashed line.} }
	\label{fig7a}
\end{figure}

\begin{figure}
	\centering
	\includegraphics[width=0.85\linewidth]{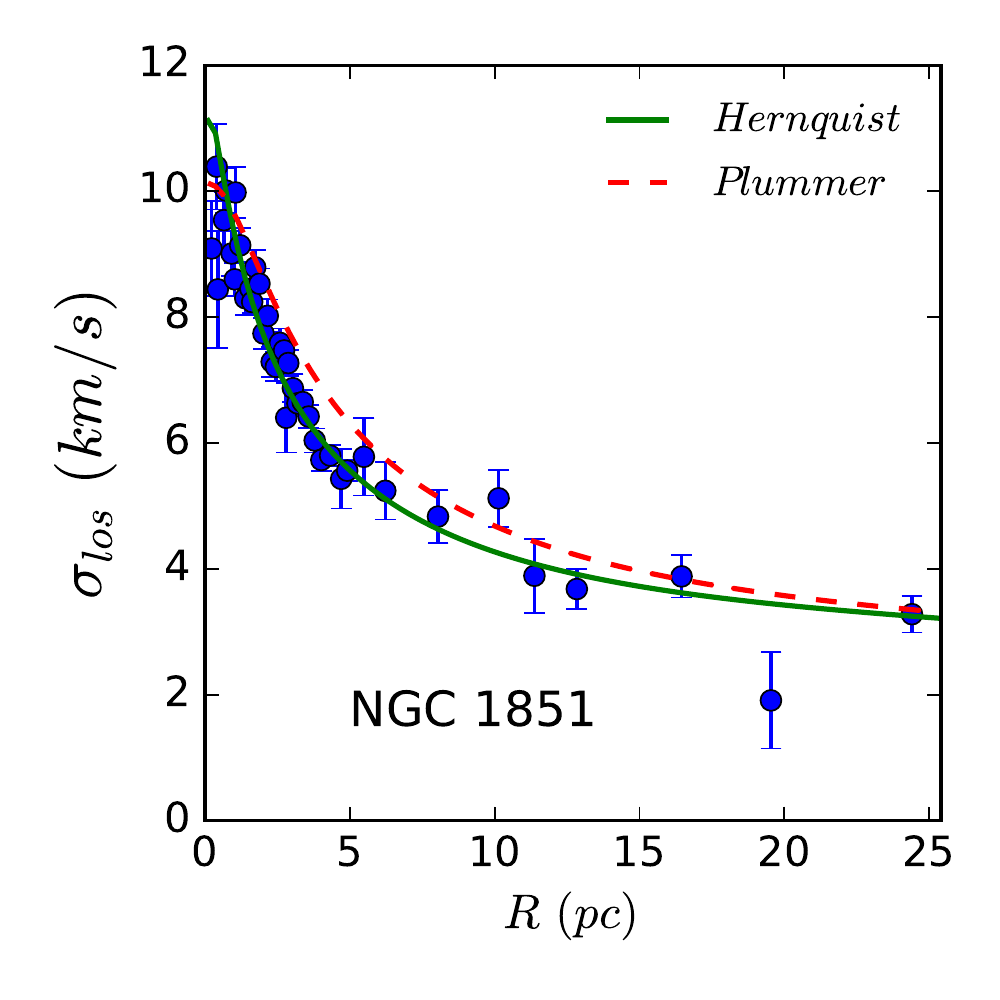}%{gc_mass}
	\caption[]{\textbf{The plot shows the measured velocity dispersion of NGC 1851 along with the best fit Weyl gravity profile with Hernquist mass model (solid green line) and the predicted Weyl gravity profile from the Plummer mass model (dashed red line).}}
	\label{fig7b} 
\end{figure}

In this paper, we approximate the GC mass profile with a simple Hernquist model. While such approximation is acceptable for the sake of simplicity, more adequate mass models, e.g. best-fit King's profile or Plummer profile to the observational density/surface brightness data, should ideally be used. It is thus worth investigating how the results would change if a more accurate mass density profile is chosen. We find that the choice of a different (and more accurate) mass model does not alter the final conclusion of the paper much. To demonstrate this, we consider NGC 1851 and choose a Plummer model, obtained by fitting surface density profile from \cite{trager1995}, to describe the mass density of the cluster.:
\begin{equation}
\rho_\mathrm{p}(r)=\frac{3M_{tot}}{\pi r^{3}_s}\Big(1+\frac{r^2}{r^2_s}\Big)^{(-5/2)},
\label{eqn22}
\end{equation}
where $M_{tot}=L(\frac{M}{L})$. The best fit values of $L$, $\frac{M}{L}$ and scale radius $r_s$ are taken from \cite{jeffreson2017gaia}. Figure \ref{fig7a} reports the two mass profiles of our interest. In the innermost region ($r<1$ pc), we notice a mismatch between the Hernquist profile and Plummer profile. However, the difference almost vanishes for larger radial distances. In Figure \ref{fig7b}, we show the reported dispersion data (blue dots) from \cite{baumgardt2018catalogue} along with the best-fit Weyl gravity profile with Hernquist mass model (green line) and the resultant Weyl gravity profile from the Plummer mass model (red dashed line). We notice a remarkable similarity between these two Weyl gravity dispersion profiles. This suggests that a best-fit mass model to the surface brightness data would be able to yield a reasonable agreement between the dispersion data and Weyl gravity predictions.

%%%%%%%%%%%%%%%%%%%%%%%%%%%%%%%%%%%%%%%%%%%%%%%%%%%%%%%%%%%%%%%%%%%%%%%%%%%%%%%%%%%%%%%%%%%%%%%%
%%%%%%%%%%%%%%%%%%%%%%%%%%%%%%%%%%%%%%%%%%%%%%%%%%%%%%%%%%%%%%%%%%%%%%%%%%%%%%%%%%%%%%%%%%%%%%%%
%%%%%%%%%%%%%%%%%%%%%%%%%%%%%%%%%%%%%%%%%%%%%%%%%%%%%%%%%%%%%%%%%%%%%%%%%%%%%%%%%%%%%%%%%%%%%%%%
%%%%%%%%%%%%%%%%%%%%%%%%%%%%%%%%%%%%%%%%%%%%%%%%%%%%%%%%%%%%%%%%%%%%%%%%%%%%%%%%%%%%%%%%%%%%%%%%

\section{Discussions and Conclusion}
\label{sec3}
In this paper, we present a phenomenological test for Weyl gravity using the observed velocity dispersion profiles for 20 GCs for which updated observational data is available. 
We have assumed the GCs to be spherically symmetric and non-rotating, and modelled them using simple Hernquist mass profile. Furthermore, we have considered the dispersion profile to be totally isotropic, have taken the mass-to-light ratio to be constant throughout the cluster and ignored any effect of tidal heating and external gravitational pull of the Milky Way on the GCs. In reality, not all of the assumptions strictly hold. For example, some of the GCs are known to rotate slowly \citep{bianchini2018internal}, the mass-to-light ratio varies radially \citep{lane2010} and the anisotropy parameter have non-zero values \citep{watkins2015hubble}. Still, we have been able to obtain excellent fits to data using Weyl gravity. It would be interesting to relax some of these assumptions and see how Weyl gravity fits change. Another interesting direction would be to use King's or Plummer mass model, whose density profile best-fit the observational surface brightness data, to fit the dispersion profiles of all the GCs considered in the paper. We leave these for future explorations.\\

Interestingly, the GCs studied in this paper are different from each other in terms of size, luminosity and distances from the galactic center; and show diverse characteristics in dispersion profiles (declining/flat) \citep{baumgardt2018catalogue}. Nonetheless, dispersion profiles for all of them could be well fitted using Weyl gravity. Weyl gravity fits to the observed dispersion profiles have resulted mass-to-light ratio ranging in between $1.0 < \frac{M}{L} < 2.4 $ (in solar unit), which is reasonable for GCs.  Moreover, total mass for the GCs inferred through  Weyl gravity fit is in agreement with the total mass derived from stellar population modelling by \cite{mclaughlin2005resolved} (Table \ref{tab2}). We have further showed that the asymptotic values of the best-fit dispersion profile and the inferred total mass for the GCs obtained via Weyl gravity fit is related through a generic Tully-Fisher-like relation, common for modified gravity theories at the galactic scale. \\

In summary, we have demonstrated that Weyl conformal gravity is consistent with the observed dynamics of a large number of globular clusters with varying luminosity, length-scale and concentration. Our work therefore suggests that Weyl conformal gravity can account for both the galaxy dynamics as well as the dynamics at the scale of globular clusters without any need to invoke dark matter.
%%%%%%%%%%%%%%%%%%%%%%%%%%%%%%%%%%%%%%%%%%%%%%%%%%%%%%%%%%%%%%%%%%%%%%%%%%%%%%%%%%%%%%%%%%%%%%%%
%%%%%%%%%%%%%%%%%%%%%%%%%%%%%%%%%%%%%%%%%%%%%%%%%%%%%%%%%%%%%%%%%%%%%%%%%%%%%%%%%%%%%%%%%%%%%%%%
\section*{Acknowledgement}
%\blindtext
I am grateful to Koushik Dutta for his encouragement throughout this work and useful comments on the manuscript. I would like to thank Rahul Kashyap, M. K. Haris and Abhishek Banerjee for vivid discussions. I am also thankful to Riccardo Scarpa for fruitful communications. I sincerely thank the anonymous referees for constructive criticism and specially for pointing out the updated velocity dispersion datasets which have been crucial to improve the scope of the paper significantly. Finally, I thank ICTS-TIFR for support through Long Term Visiting Students Program. 
%%%%%%%%%%%%%%%%%%%%%%%%%%%%%%%%%%%%%%%%%%%%%%%%%%%%%%%%%%%%%%%%%%%%%%%%%%%%%%%%%%%%%%%%%%%%%%%%
%%%%%%%%%%%%%%%%%%%%%%%%%%%%%%%%%%%%%%%%%%%%%%%%%%%%%%%%%%%%%%%%%%%%%%%%%%%%%%%%%%%%%%%%%%%%%%%%
%\bibliography{Tousif_Weyl_2019}

\begin{thebibliography}{}
	\makeatletter
	\relax
	\def\mn@urlcharsother{\let\do\@makeother \do\$\do\&\do\#\do\^\do\_\do\%\do\~}
	\def\mn@doi{\begingroup\mn@urlcharsother \@ifnextchar [ {\mn@doi@}
		{\mn@doi@[]}}
	\def\mn@doi@[#1]#2{\def\@tempa{#1}\ifx\@tempa\@empty \href
		{http://dx.doi.org/#2} {doi:#2}\else \href {http://dx.doi.org/#2} {#1}\fi
		\endgroup}
	\def\mn@eprint#1#2{\mn@eprint@#1:#2::\@nil}
	\def\mn@eprint@arXiv#1{\href {http://arxiv.org/abs/#1} {{\tt arXiv:#1}}}
	\def\mn@eprint@dblp#1{\href {http://dblp.uni-trier.de/rec/bibtex/#1.xml}
		{dblp:#1}}
	\def\mn@eprint@#1:#2:#3:#4\@nil{\def\@tempa {#1}\def\@tempb {#2}\def\@tempc
		{#3}\ifx \@tempc \@empty \let \@tempc \@tempb \let \@tempb \@tempa \fi \ifx
		\@tempb \@empty \def\@tempb {arXiv}\fi \@ifundefined
		{mn@eprint@\@tempb}{\@tempb:\@tempc}{\expandafter \expandafter \csname
			mn@eprint@\@tempb\endcsname \expandafter{\@tempc}}}
	
	\bibitem[\protect\citeauthoryear{Baumgardt}{Baumgardt}{2016}]{baumgardt2016n}
	Baumgardt H.,  2016, Monthly Notices of the Royal Astronomical Society, 464,
	2174
	
	\bibitem[\protect\citeauthoryear{Baumgardt \& Hilker}{Baumgardt \&
		Hilker}{2018}]{baumgardt2018catalogue}
	Baumgardt H.,  Hilker M.,  2018, Monthly Notices of the Royal Astronomical
	Society, 478, 1520
	
	\bibitem[\protect\citeauthoryear{Baumgardt, Hilker, Sollima  \&
		Bellini}{Baumgardt et~al.}{2018}]{baumgardt2018mean}
	Baumgardt H.,  Hilker M.,  Sollima A.,   Bellini A.,  2018, Monthly Notices of
	the Royal Astronomical Society, 482, 5138
	
	\bibitem[\protect\citeauthoryear{Bender \& Mannheim}{Bender \&
		Mannheim}{2008}]{bender2008no}
	Bender C.~M.,  Mannheim P.~D.,  2008, Physical Review Letters, 100, 110402
	
	\bibitem[\protect\citeauthoryear{Bertone, Hooper  \& Silk}{Bertone
		et~al.}{2005}]{bertone2005particle}
	Bertone G.,  Hooper D.,   Silk J.,  2005, Physics Reports, 405, 279
	
	\bibitem[\protect\citeauthoryear{Bianchini, van~der Marel, del Pino, Watkins,
		Bellini, Fardal, Libralato  \& Sills}{Bianchini
		et~al.}{2018}]{bianchini2018internal}
	Bianchini P.,  van~der Marel R.,  del Pino A.,  Watkins L.,  Bellini A.,
	Fardal M.,  Libralato M.,   Sills A.,  2018, Monthly Notices of the Royal
	Astronomical Society, 481, 2125
	
	\bibitem[\protect\citeauthoryear{Binney \& Tremaine}{Binney \&
		Tremaine}{1987}]{bt1987}
	Binney J.,  Tremaine S.,  1987, Galactic Dynamics, Princeton Univ
	
	\bibitem[\protect\citeauthoryear{Bressan, Marigo, Girardi, Salasnich, Dal~Cero,
		Rubele  \& Nanni}{Bressan et~al.}{2012}]{bressan2012}
	Bressan A.,  Marigo P.,  Girardi L.,  Salasnich B.,  Dal~Cero C.,  Rubele S.,
	Nanni A.,  2012, Monthly Notices of the Royal Astronomical Society, 427, 127
	
	\bibitem[\protect\citeauthoryear{Clowe, Brada{\v{c}}, Gonzalez, Markevitch,
		Randall, Jones  \& Zaritsky}{Clowe et~al.}{2006}]{clowe2006direct}
	Clowe D.,  Brada{\v{c}} M.,  Gonzalez A.~H.,  Markevitch M.,  Randall S.~W.,
	Jones C.,   Zaritsky D.,  2006, The Astrophysical Journal Letters, 648, L109
	
	\bibitem[\protect\citeauthoryear{Cutajar \& Zarb~Adami}{Cutajar \&
		Zarb~Adami}{2014}]{weylcluster3}
	Cutajar D.,  Zarb~Adami K.,  2014, Monthly Notices of the Royal Astronomical
	Society, 441, 1291
	
	\bibitem[\protect\citeauthoryear{Diaferio \& Ostorero}{Diaferio \&
		Ostorero}{2009}]{weylcluster1}
	Diaferio A.,  Ostorero L.,  2009, Monthly Notices of the Royal Astronomical
	Society, 393, 215
	
	\bibitem[\protect\citeauthoryear{Drukier, Cohn, Lugger, Slavin, Berrington  \&
		Murphy}{Drukier et~al.}{2007}]{drukier2007global}
	Drukier G.,  Cohn H.,  Lugger P.,  Slavin S.,  Berrington R.,   Murphy B.,
	2007, The Astronomical Journal, 133, 1041
	
	\bibitem[\protect\citeauthoryear{Dutta \& Islam}{Dutta \& Islam}{2018}]{kt2018}
	Dutta K.,  Islam T.,  2018, \mn@doi [Phys. Rev. D]
	{10.1103/PhysRevD.98.124012}, 98, 124012
	
	\bibitem[\protect\citeauthoryear{Haghi, Baumgardt, Kroupa, Grebel, Hilker  \&
		Jordi}{Haghi et~al.}{2009}]{haghi2009testing}
	Haghi H.,  Baumgardt H.,  Kroupa P.,  Grebel E.~K.,  Hilker M.,   Jordi K.,
	2009, Monthly Notices of the Royal Astronomical Society, 395, 1549
	
	\bibitem[\protect\citeauthoryear{Haghi, Baumgardt  \& Kroupa}{Haghi
		et~al.}{2011}]{haghi2011distant}
	Haghi H.,  Baumgardt H.,   Kroupa P.,  2011, Astronomy \& Astrophysics, 527,
	A33
	
	\bibitem[\protect\citeauthoryear{Hernandez, Jim{\'e}nez  \& Allen}{Hernandez
		et~al.}{2012}]{gctidal}
	Hernandez X.,  Jim{\'e}nez M.,   Allen C.,  2012, Monthly Notices of the Royal
	Astronomical Society, 428, 3196
	
	\bibitem[\protect\citeauthoryear{Hernquist}{Hernquist}{1990}]{hern}
	Hernquist L.,  1990, The Astrophysical Journal, 356, 359
	
	\bibitem[\protect\citeauthoryear{Horne}{Horne}{2006}]{weylcluster2}
	Horne K.,  2006, Monthly Notices of the Royal Astronomical Society, 369, 1667
	
	\bibitem[\protect\citeauthoryear{Jeffreson et~al.,}{Jeffreson
		et~al.}{2017}]{jeffreson2017gaia}
	Jeffreson S.~M.,  et~al., 2017, Monthly Notices of the Royal Astronomical
	Society, 469, 4740
	
	\bibitem[\protect\citeauthoryear{Kamann et~al.,}{Kamann
		et~al.}{2017}]{kamann2017stellar}
	Kamann S.,  et~al., 2017, Monthly Notices of the Royal Astronomical Society,
	473, 5591
	
	\bibitem[\protect\citeauthoryear{Kennedy}{Kennedy}{2014}]{kennedy2014application}
	Kennedy G.~F.,  2014, Monthly Notices of the Royal Astronomical Society, 444,
	3328
	
	\bibitem[\protect\citeauthoryear{King}{King}{1966}]{king1966structure}
	King I.~R.,  1966, The Astronomical Journal, 71, 64
	
	\bibitem[\protect\citeauthoryear{Kruijssen \& Mieske}{Kruijssen \&
		Mieske}{2009}]{kruijssen2009}
	Kruijssen J.~D.,  Mieske S.,  2009, Astronomy \& Astrophysics, 500, 785
	
	\bibitem[\protect\citeauthoryear{K{\"u}pper, Kroupa, Baumgardt  \&
		Heggie}{K{\"u}pper et~al.}{2010}]{kupper2010}
	K{\"u}pper A. H.~W.,  Kroupa P.,  Baumgardt H.,   Heggie D.~C.,  2010, Monthly
	Notices of the Royal Astronomical Society, 407, 2241
	
	\bibitem[\protect\citeauthoryear{Lane et~al.,}{Lane et~al.}{2010}]{lane2010}
	Lane R.~R.,  et~al., 2010, Monthly Notices of the Royal Astronomical Society,
	406, 2732
	
	\bibitem[\protect\citeauthoryear{Mannheim}{Mannheim}{1997}]{weylrot2}
	Mannheim P.~D.,  1997, The Astrophysical Journal, 479, 659
	
	\bibitem[\protect\citeauthoryear{Mannheim}{Mannheim}{2006}]{weylrot5}
	Mannheim P.~D.,  2006, Progress in Particle and Nuclear Physics, 56, 340
	
	\bibitem[\protect\citeauthoryear{Mannheim}{Mannheim}{2007}]{mannheim2007schwarzschild}
	Mannheim P.~D.,  2007, Physical Review D, 75, 124006
	
	\bibitem[\protect\citeauthoryear{Mannheim \& Kazanas}{Mannheim \&
		Kazanas}{1989}]{weyl1}
	Mannheim P.~D.,  Kazanas D.,  1989, The Astrophysical Journal, 342, 635
	
	\bibitem[\protect\citeauthoryear{Mannheim \& O’Brien}{Mannheim \&
		O’Brien}{2011}]{weylrot3}
	Mannheim P.~D.,  O’Brien J.~G.,  2011, Physical review letters, 106, 121101
	
	\bibitem[\protect\citeauthoryear{Mannheim \& O’Brien}{Mannheim \&
		O’Brien}{2012}]{weylrot1}
	Mannheim P.~D.,  O’Brien J.~G.,  2012, Physical Review D, 85, 124020
	
	\bibitem[\protect\citeauthoryear{McLaughlin \& van~der Marel}{McLaughlin \&
		van~der Marel}{2005}]{mclaughlin2005resolved}
	McLaughlin D.~E.,  van~der Marel R.~P.,  2005, The Astrophysical Journal
	Supplement Series, 161, 304
	
	\bibitem[\protect\citeauthoryear{Moffat \& Toth}{Moffat \&
		Toth}{2008}]{moffattoth}
	Moffat J.,  Toth V.,  2008, The Astrophysical Journal, 680, 1158
	
	\bibitem[\protect\citeauthoryear{Moore}{Moore}{1996}]{gcdm2}
	Moore B.,  1996, The Astrophysical Journal Letters, 461, L13
	
	\bibitem[\protect\citeauthoryear{O'Brien \& Moss}{O'Brien \&
		Moss}{2015}]{obrien}
	O'Brien J.~G.,  Moss R.~J.,  2015, in Journal of Physics: Conference Series. p.
	012002
	
	\bibitem[\protect\citeauthoryear{O’Brien \& Mannheim}{O’Brien \&
		Mannheim}{2012}]{weylrot4}
	O’Brien J.~G.,  Mannheim P.~D.,  2012, Monthly Notices of the Royal
	Astronomical Society, 421, 1273
	
	\bibitem[\protect\citeauthoryear{Peebles \& Ratra}{Peebles \&
		Ratra}{2003}]{peebles2003cosmological}
	Peebles P. J.~E.,  Ratra B.,  2003, Reviews of modern physics, 75, 559
	
	\bibitem[\protect\citeauthoryear{Phinney}{Phinney}{1993}]{gcdm1}
	Phinney E.,  1993, in Structure and Dynamics of Globular Clusters. p.~141
	
	\bibitem[\protect\citeauthoryear{Plummer}{Plummer}{1911}]{plummer1911problem}
	Plummer H.~C.,  1911, Monthly notices of the royal astronomical society, 71,
	460
	
	\bibitem[\protect\citeauthoryear{Scarpa, Marconi  \& Gilmozzi}{Scarpa
		et~al.}{2003}]{scarpa2003using}
	Scarpa R.,  Marconi G.,   Gilmozzi R.,  2003, Astronomy \& Astrophysics, 405,
	L15
	
	\bibitem[\protect\citeauthoryear{Scarpa, Marconi  \& Gilmozzi}{Scarpa
		et~al.}{2004a}]{scarpa2004using}
	Scarpa R.,  Marconi G.,   Gilmozzi R.,  2004a, arXiv preprint astro-ph/0411078
	
	\bibitem[\protect\citeauthoryear{Scarpa, Marconi  \& Gilmozzi}{Scarpa
		et~al.}{2004b}]{ngcothers2}
	Scarpa R.,  Marconi G.,   Gilmozzi R.,  2004b, in Symposium-International
	Astronomical Union. pp 215--216
	
	\bibitem[\protect\citeauthoryear{Scarpa, Marconi, Gilmozzi  \& Carraro}{Scarpa
		et~al.}{2007}]{scarpa2007using}
	Scarpa R.,  Marconi G.,  Gilmozzi R.,   Carraro G.,  2007, Astronomy \&
	Astrophysics, 462, L9
	
	\bibitem[\protect\citeauthoryear{Scarpa, Marconi, Carraro, Falomo  \&
		Villanova}{Scarpa et~al.}{2011}]{ngc1851a1904}
	Scarpa R.,  Marconi G.,  Carraro G.,  Falomo R.,   Villanova S.,  2011,
	Astronomy \& Astrophysics, 525, A148
	
	\bibitem[\protect\citeauthoryear{Sultana, Kazanas  \& Said}{Sultana
		et~al.}{2012}]{perihelion}
	Sultana J.,  Kazanas D.,   Said J.~L.,  2012, Physical Review D, 86, 084008
	
	\bibitem[\protect\citeauthoryear{Trager, King  \& Djorgovski}{Trager
		et~al.}{1995}]{trager1995}
	Trager S.,  King I.~R.,   Djorgovski S.,  1995, The Astronomical Journal, 109,
	218
	
	\bibitem[\protect\citeauthoryear{Watkins, van~der Marel, Bellini  \&
		Anderson}{Watkins et~al.}{2015}]{watkins2015hubble}
	Watkins L.~L.,  van~der Marel R.~P.,  Bellini A.,   Anderson J.,  2015, The
	Astrophysical Journal, 803, 29
	
	\bibitem[\protect\citeauthoryear{Weyl}{Weyl}{1918}]{weyl1918}
	Weyl H.,  1918, Mathematische Zeitschrift, 2, 384
	
	\makeatother
\end{thebibliography}

\label{lastpage}
\end{document}